\documentclass[
reprint,
superscriptaddress,
 amsmath,amssymb,
 aps,
 footinbib,
floatfix
]{revtex4-2}
\usepackage{float}
\usepackage{graphicx}
\usepackage{bm}
\usepackage{hyperref}
\usepackage{physics}
\usepackage{xcolor}

\newcommand{\wide}[1]{\resizebox{\linewidth}{!}{$\displaystyle #1$}}
\newcommand{\VHD}{V^{HD}}
\newcommand{\xme}[1]{m_{x1}}
\newcommand{\pme}[1]{m_{p1}}
\begin{document}

\title{Quantum teleportation enhanced optical displacement sensing}

\newcommand{\addrbigQ}{Center for Macroscopic Quantum States bigQ, Department of Physics, Technical University of Denmark, Fysikvej 307, DK-2800 Kgs.\ Lyngby, Denmark}

\author{Mateusz Kicinski}
\affiliation{\addrbigQ}
\author{Clémentine Rouvière}
\affiliation{\addrbigQ}
\author{Benjamin Lundgren Larsen}
\affiliation{\addrbigQ}
\author{Jonas S. Neergaard-Nielsen}
\affiliation{\addrbigQ}
\author{Ulrik L. Andersen}
\affiliation{\addrbigQ}

\begin{abstract}
New measurement techniques are continuously being developed in order to
explore new phenomena and learn more about those already known. Quantum
metrology provides tools that allow us to surpass the limits on the precision
of measurements set by classical physics. Another route to higher
precision is the multipass protocol, which is, however, difficult to realize
and of limited practical use. Combined with quantum resources, it can be
implemented by utilizing quantum teleportation. Here we experimentally demonstrate multipass
displacement sensing in which the repeated interactions are realized by
quantum teleportation through a time-multiplexed continuous-variable cluster
state. In our protocol, the information about the displacement is teleported
between temporal modes that are entangled in the time domain, and in this way
accumulates over up to five consecutive teleportations. The sensitivity
reached surpasses the classical limit of repeated single-pass measurements by
up to $(13.4 \pm 2.2)\%$. The results agree with the theoretical model that
we derive.
\end{abstract}

\maketitle

\section{Introduction}
\label{sec: itroduction}

Metrology is one of the most fundamental branches of science, and its impact
reaches far beyond research. It provides the tools to observe what has never
been seen before, to measure phenomena hidden from the naked eye, and to study
objects ranging from single particles to systems as complex as living
organisms. The development of measurement techniques is a never-ending process
that constantly leads to discoveries across many fields. Within the last
decade alone, it has enabled the detection of gravitational
waves~\cite{Abbott2016}, to create the first image of a black hole~\cite{EHT2019} or perform
imaging of proteins at atomic resolution~\cite{Nakane2020}.

Quantum physics pushes sensing beyond what is allowed by classical physics ~\cite{Giovannetti2004,Giovannetti2011,Degen2017}. Various quantum sensing platforms are continuously under development, with notable examples including nitrogen-vacancy centers in diamond~\cite{Taylor2008,Barry2020}, silicon-vacancy centers~\cite{Nguyen2018,Simin2016} or radio-frequency and microwave sensors based on Rydberg atoms and optomechanical
transducers~\cite{Sedlacek2012,Bagci2014}. Among these techniques, optical interferometry is
particularly useful. The reason behind is that light couples to a wide range of physical
quantities and is less susceptible to environmental noise. Its
applications range from anything like gravitational-wave detection~\cite{Abbott2016} up to
biomedical imaging~\cite{Taylor2016}.

Classical physics sets the limit on precision of optical measurements. This limit is called the standard quantum limit (SQL), where the uncertainty of measurements scales with numbers of information carriers (photons). Quantum mechanics allows to surpass this limit and sets limitation in precision way beyond what is classically achievable ~\cite{Holland1993,Giovannetti2006,Demkowicz2012}. Several approaches have
been explored to bring quantum resources into sensing, including squeezed
light~\cite{Caves1981,Xiao1987,LIGO2011,Schnabel2017, JensBeyondNOON, VQAJens2025} , NOON
states ~\cite{Mitchell2004,Dowling2008,Slussarenko2017} or distributed
sensing technique~\cite{Zhuang2018,Guo2020,Zhang2021}. A
particularly interesting and less explored technique is the multipass
protocol, in which the probe interacts with the sample several times so that
the signal accumulates
coherently~\cite{Giovannetti2006,Juffmann2016}. Its direct
optical implementation, however, is technologically demanding, which limits
its practicality. Quantum teleportation offers a way around this problem:
instead of physically redirecting the probe, its state can be repeatedly
teleported back to interact with the sample repeatedly ~\cite{Johanness}. Such entanglement is the building block of continuous-variable cluster states, the resource for measurement-based quantum computing ~\cite{Raussendorf2001,Menicucci2006,Yokoyama2013,Larsen2019, AbhiScalable}.

In this work, we experimentally realize the teleportation-based multipass protocol proposed in Ref.~\cite{Johanness}, adjusted from phase to optical displacement sensing. The protocol consecutively teleports the information about the optical displacement that our probe state undergoes, and thereby reaches sensitivity that cannot be achieved by repeating a single-pass measurement. We experimentally show that the quantum resources allow the protocol to surpass the classical limit, and provide a theoretical and numerical model of the experiment. Our approach works entirely in the continuous-variable regime, relies on well-established resources, and uses the vacuum state as the input. Although the experiment is challenging to build, it combines only well-understood techniques and can therefore be implemented in many quantum optics laboratories.

\section{Experimental principle}
\label{sec:principle}

\begin{figure*}[t]
  \centering
  \includegraphics[width=\textwidth]{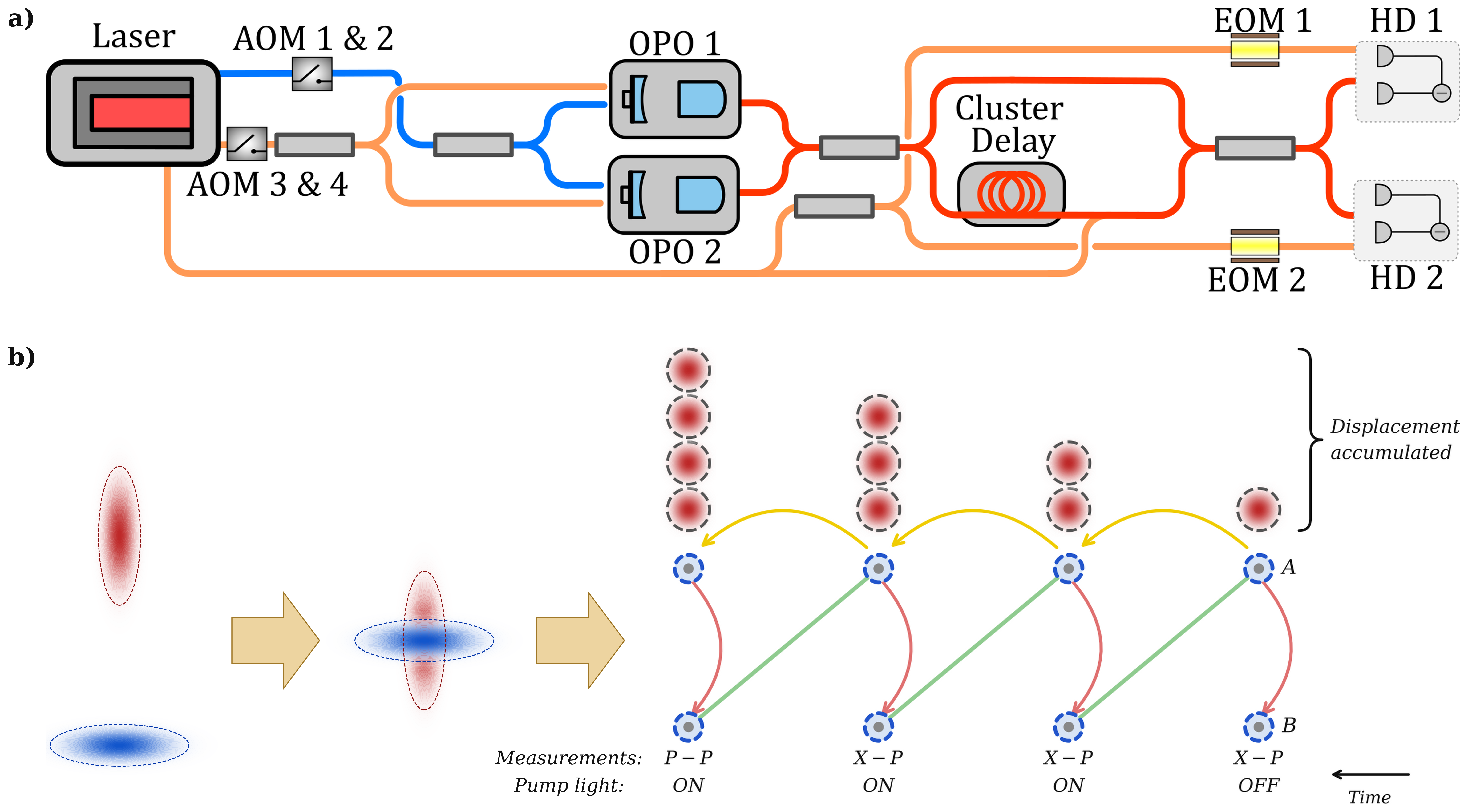}
  \caption{Principle of the experiment. (a) Simplified experimental scheme.
  Blue: 775nm pump light; orange: 1550nm probe beam, local oscillators and displacement beam; red:
  1550nm squeezed light propagating alongside gated probe beam. AOM: acousto-optic modulator; OPO: optical parametric
  oscillator; EOM: electro-optic phase modulator; HD: homodyne detector.
  (b) Evolution of the state. Left: the two squeezed vacuum states, squeezed
  in $\hat{x}$ (red) and in $\hat{p}$ (blue). Centre: their interference on
  the first coupler gives an EPR state. Right: cluster state structure for
  four temporal modes, with time running from right to left. $A$ and $B$ are
  the delayed and the undelayed mode of each temporal mode. Red arrows
  connect modes generated together, green lines connect modes that are
  measured together, and yellow arrows mark the teleportation of the state
  to the next temporal mode. The stacked circles above mode $A$ show the
  displacement accumulated up to that mode. The labels give the quadratures
  measured by the two homodyne detectors and the state of the pump.}
  \label{fig:protocol}
\end{figure*}

The protocol follows the multi-pass teleportation scheme of
Ref.~\cite{Johanness}, with the phase shift replaced by a displacement.
The difference between displacement and phase sensing is described in
Section~\ref{sec: disp_vs_phase}. The experimental scheme is shown in Fig.~\ref{fig:protocol}a and the
evolution of the optical state in Fig.~\ref{fig:protocol}b. The 775~nm pump
light, gated by AOM~1~\&~2, is
sent into two optical parametric oscillators (OPO~1 and OPO~2) to initiate
parametric down-conversion. As a result, each OPO emits squeezed vacuum split into temporal bins of $T = 308$~ns, where $1/T$ is well within the
bandwidth of the OPOs. The two states are squeezed in orthogonal
quadratures, one in $\hat{x}$ and the other in $\hat{p}$
(Fig.~\ref{fig:protocol}b, left). This is ensured by setting the relative
phase between the pump and the probe light in each OPO
appropriately. Interference of the squeezed light on a 50:50 coupler generates an Einstein--Podolsky--Rosen (EPR) state distributed
over the two coupler outputs (Fig.~\ref{fig:protocol}b, centre). One of the
outputs is delayed by exactly one temporal mode (Cluster Delay) and then
interferes on a second 50:50 coupler with the undelayed mode of the following
EPR pair. Consecutive EPR pairs are in this way linked into the
one-dimensional cluster-state-like structure shown on the right of
Fig.~\ref{fig:protocol}b, in which each pair acts as a teleportation channel
from one temporal mode to the next.

The displacement to be sensed is applied by a weak coherent
beam, derived from the same light source and modulated at approximately 6.49~MHz,
which is injected into the delayed arm after the delay line. Every temporal
mode is therefore displaced after the delay line by the same amount at the
6.49~MHz sideband. The two outputs of the second coupler are measured by two
homodyne detectors (HD~1 and HD~2). Electro-optic modulators in the
local-oscillator paths (EOM~1 and EOM~2) set the quadrature measured by each
detector and switch it from one temporal mode to the next.

A probe beam, gated by AOM~3~\&~4, seeds both OPOs and provides light used as
a phase reference for all locks. The experiment alternates between a 5~ms
lock period, in which the probe is on and the relative phases are stabilized,
and a 5~ms measurement period, in which the probe is off, the controllers
hold their last output and the data is collected. The complete setup is
described in Section~\ref{sec: full_experimental_setup}.

The right-hand side of Fig.~\ref{fig:protocol}b visualizes how the protocol
utilizes the cluster state structure. Each column is one temporal mode, and
time runs from right to left. In every temporal mode, $A$ denotes the mode
that passes through the delay line and $B$ represents the undelayed mode. A
red (dark grey) arrow connects the two modes generated in the same temporal
mode, which form an EPR pair whenever the pump is on. A green (grey) line
connects mode $A$ to mode $B$ of the following temporal mode and represents
the entanglement in the time domain caused by interference on the second
coupler. Modes created this way are measured jointly by HD~1 and HD~2. The
measurement frequency is set in post-processing by shaping the signal with
filter functions (mode functions), described in Section~\ref{sec: modes}.
The labels below each column show the quadratures measured in this joint
measurement and the state of the pump.

The protocol starts in the rightmost temporal mode, for which the pump is switched off, so that both modes are in the vacuum state and mode $A$ provides the vacuum input state. Its
displacement after the delay line is shown as a single circle above it. In an
$X$--$P$ measurement, the detectors measure the $\hat{x}$ quadrature of one
coupler output and the $\hat{p}$ quadrature of the other. This teleports the
information about the displacement onto mode $A$ of the following temporal mode (yellow arrow),
where it is displaced again. After $M-1$ sequential teleportations implemented on successive temporal modes the state has therefore
accumulated $M$ displacements, as shown by the stacked circles. In contrast
to Ref.~\cite{Johanness}, the outcome-dependent displacement that
accompanies each teleportation, and is introduced by measuring out one of two entangled modes of the cluster state structure, is not corrected by a feedforward loop. The
measurement outcomes are recorded and taken into account in post-processing.
In the last temporal mode, both detectors measure the $\hat{p}$ quadrature
($P$--$P$), which reads out the accumulated displacement and completes the
protocol. Figure~\ref{fig:protocol}b shows the case of four temporal modes,
that is, three teleportation steps and four displacements.

\section{Results}
\label{sec: results}

\begin{figure}[!tp]
  \centering
  \makebox[\linewidth][l]{\textbf{(a)}}\\[-0.5ex]
  \includegraphics[width=0.94\linewidth]{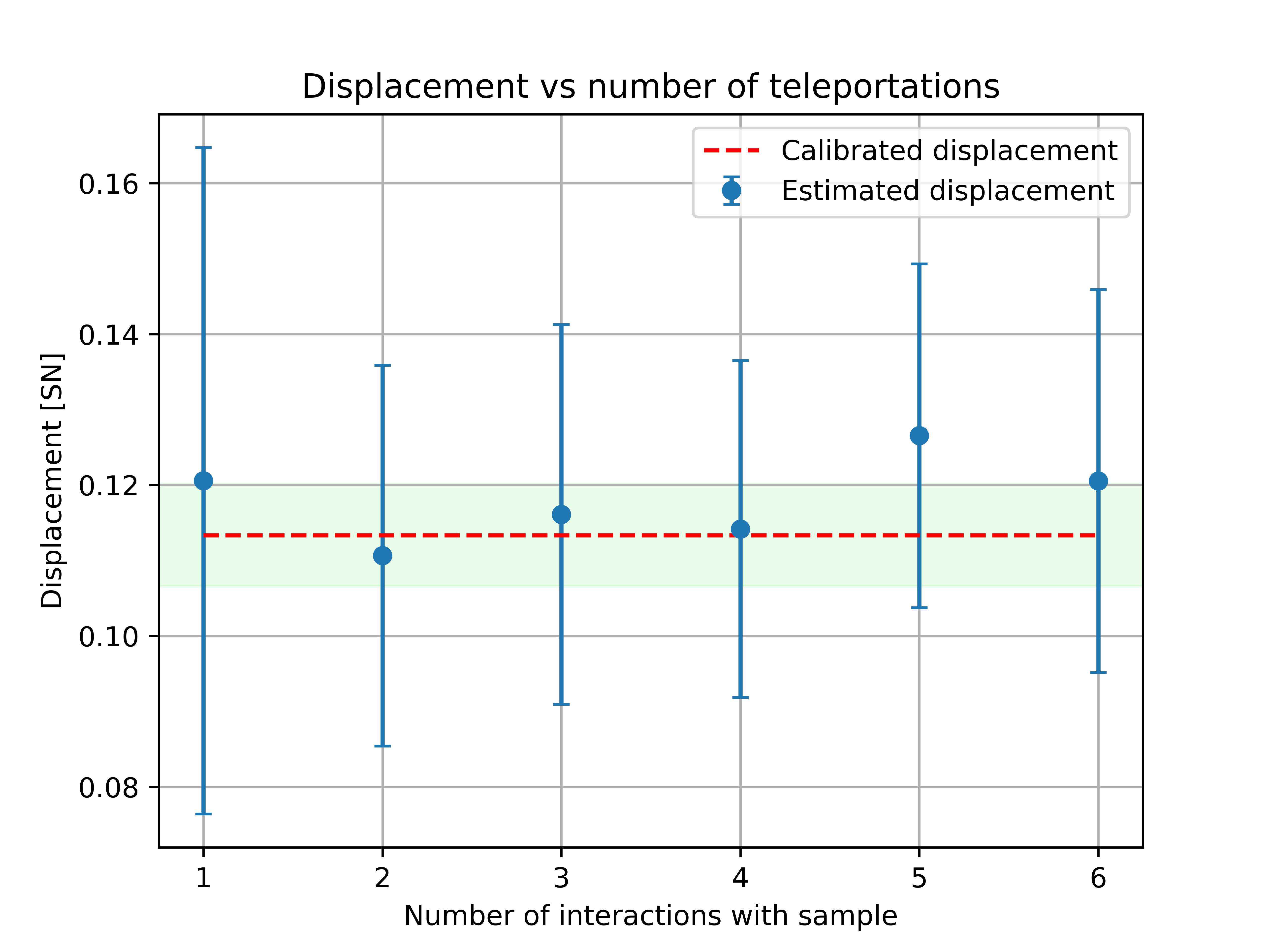}\\[0.5ex]
  \makebox[\linewidth][l]{\textbf{(b)}}\\[-0.5ex]
  \includegraphics[width=0.94\linewidth]{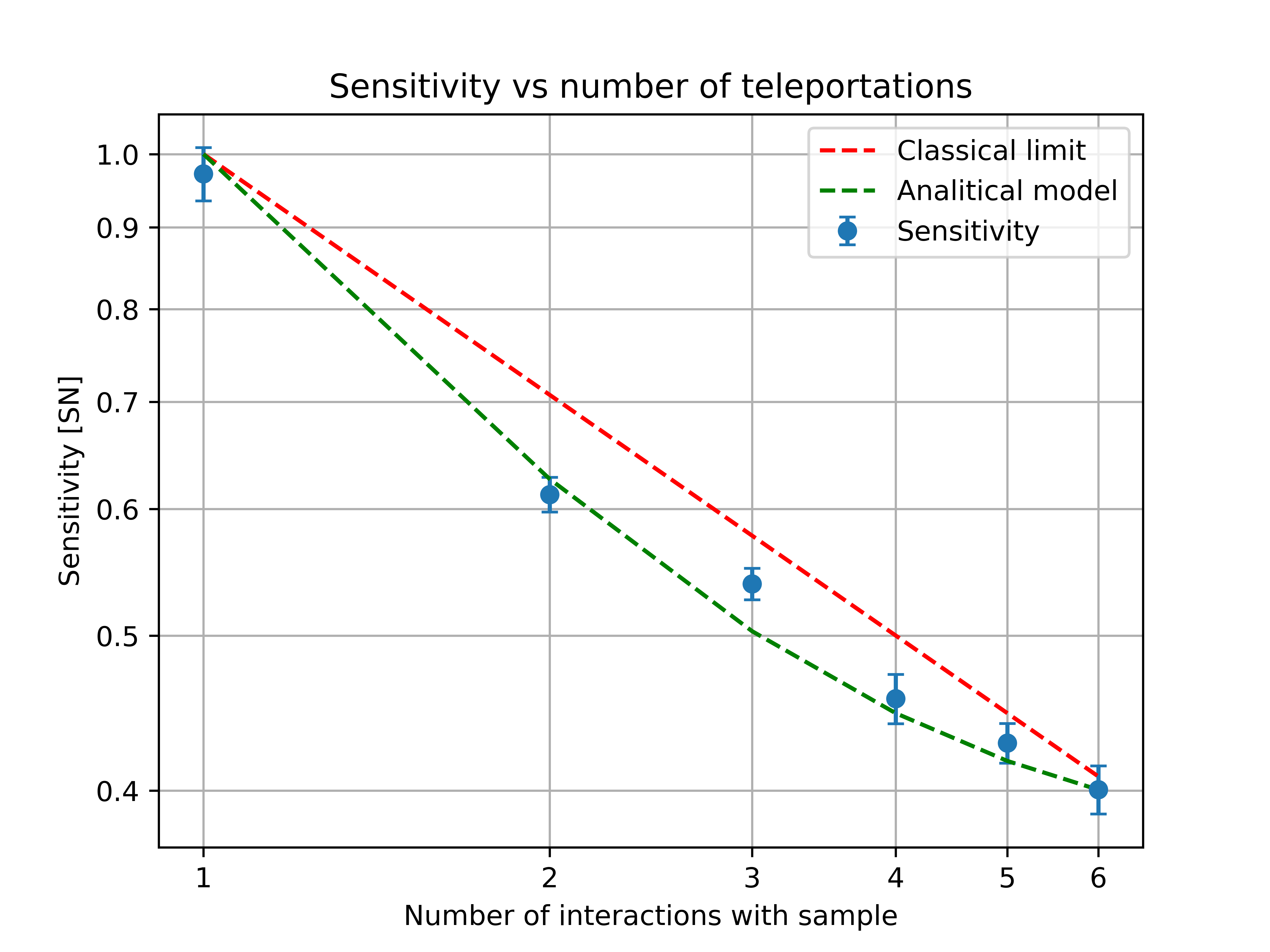}\\[0.5ex]
  \makebox[\linewidth][l]{\textbf{(c)}}\\[-0.5ex]
  \includegraphics[width=0.94\linewidth]{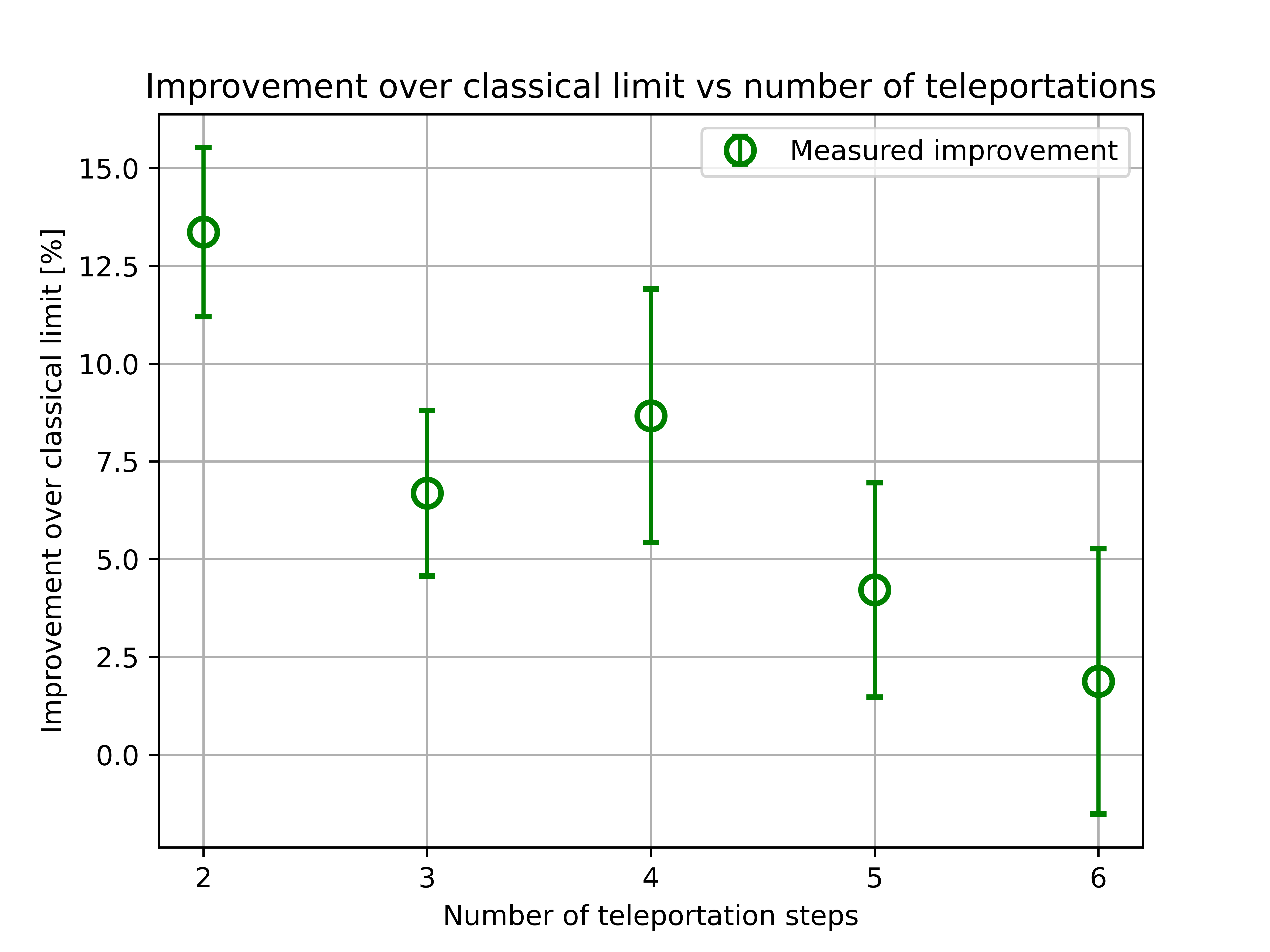}
  \caption{Results as a function of the number of interactions with the
  sample, $M$. (a) Estimated displacement. Red dashed line and shaded band:
  calibrated displacement and its uncertainty. (b) Sensitivity of the
  displacement estimate on logarithmic axes. Red dashed line: classical
  limit, $1/\sqrt{M}$. Green dashed line: analytical model. (c) Improvement
  of the measured sensitivity over the classical limit for the same number of
  interactions.}
  \label{fig:results}
\end{figure}

The experiment was performed for $M = 1$ to $6$ interactions with the
introduced displacement, which corresponds to up to five consecutive teleportations. A
single interaction ($M = 1$) is equivalent to a single-pass protocol. Each data point is the average of the 10 acquisitions with the lowest variance, out of about 35 recorded, and the error bars show the standard deviation of these 10 values. Each measurement is itself
obtained from the statistics of 450 runs of the experiment, for which number the results of measurements were converging to normal distribution. The results are
shown in Fig.~\ref{fig:results}.

On Figure~\ref{fig:results}a we compare value of the estimated displacement for each $M$ with the value obtained during independently performed calibration done before performing experiments. All estimates agree with the calibrated value within their uncertainties. The estimates show no trend with $M$. The estimator is therefore unbiased.

Figure~\ref{fig:results}b shows the sensitivity, i.e.\ the uncertainty of the displacement estimate, as a function of number of passes through the sample. The classical limit (standard quantum limit, SQL) is the sensitivity of $M$ independent single-pass measurements with a (vacuum) probe, which is $1/\sqrt{M}$. With teleportation, the sensitivity improves steadily with the number of interactions and for number of interactions within $2$ and $5$ (number of teleportatinons $1-4$) lies below the classical limit. For 5 teleportations average value is below the classical limit. However, with uncertainty of the measurement it crosses the SQL, thus no improvement over classical limit can be claimed. All of the measurement data points, but for number of passes through the sample equal $M=3$, are well aligned with the theoretical model that can be found in section \ref{sec:analytical_model}. Model accounts for the finite squeezing, linear optical losses and phase noise of phase locking loops in the setup. Reason for data for $M=3$ not agreeing with simulations is most probably not perfect stability of the experimental setup, its phase locking loops, pump and probe power fluctuations.

Figure~\ref{fig:results}c shows the same data as the relative improvement over the classical limit. The improvement is largest for a single teleportation step ($M = 2$), where it reaches $(13.4 \pm 2.2)\%$ and drops down to $(1.9 \pm 3.4)\%$ at $M = 6$, which is consistent with no improvement over sensing with classical protocols. This decrease is expected. Finite squeezing and imperfection in form of linear losses and phase noise are introducing gaussian noise during teleportation. This results in decreased quality of teleportation, thus indirectly improvement in sensitivity of measurements decaying over number of teleportations. For the present parameters, the model predicts that the advantage vanishes at about $M = 6$. With higher squeezing and lower losses the advantage would persist to larger $M$, and the sensitivity would approach the $1/M$ scaling of the ideal protocol~\cite{Johanness}.

\section{Discussion and outlook}

We have demonstrated experimentally a quantum teleportation based multipass optical displacement sensing protocol. Resource for performing quantum teleportation was continuous-variable optical cluster state. For one to four number of teleportations the measured sensitivity lies below the classical limit of repeated single-pass measurements, with the largest improvement of $(13.4 \pm 2.2)\%$ for a single teleportation step, while the estimated displacement remains unbiased throughout the whole experiment. This shows that the information about the displacement is transferred coherently from one temporal mode to the next and accumulates over consecutive teleportations.

The advantage over the classical limit decreases with the number of teleportations and is no longer significant for five teleportations. This behaviour is reproduced by our model and originates from the finite squeezing of the EPR states, together with optical losses and phase noise, which make every teleportation imperfect. The imperfection accumulates with each step, whereas for single pass protocol sensitivity keeps improving with the number of repetitions without additional sources of noise. The performance is therefore limited by the quality of the entanglement and not by the protocol itself, and stronger squeezing and lower losses would extend the advantage to a larger number of interactions.

Our implementation differs from the original theoretical proposal~\cite{Johanness}. The outcome-dependent displacements that accompany each teleportation are not corrected by feedforward but are taken into account in post-processing which enforces change of quantity being measured from phase shift to optical displacement (Section~\ref{sec: disp_vs_phase}). On the contrary it reduces complexity of experiment by removing need of ultrafast feedforward loop.

Several extensions of this work follow naturally. Firstly ultrafast feedforward loop for correcting measurement outcome-based displacements can be implemented in order to perform phase sensing, secondly additional feedback loop on local oscillator path can be added in order to perform adaptive measurements and last but not least the injected displacement
can be replaced by a physical sample placed in the delayed arm and in combination with correction of outcome-based displacement the setup can be brought closer to practical applications.

\bibliography{Ref_tele_sensing}

\clearpage
\onecolumngrid
\section{Methods}
\label{sec: methods}

\subsection{Full experimental System}
\label{sec: full_experimental_setup}

\begin{figure*}[t]
  \centering
  \includegraphics[width=\textwidth]{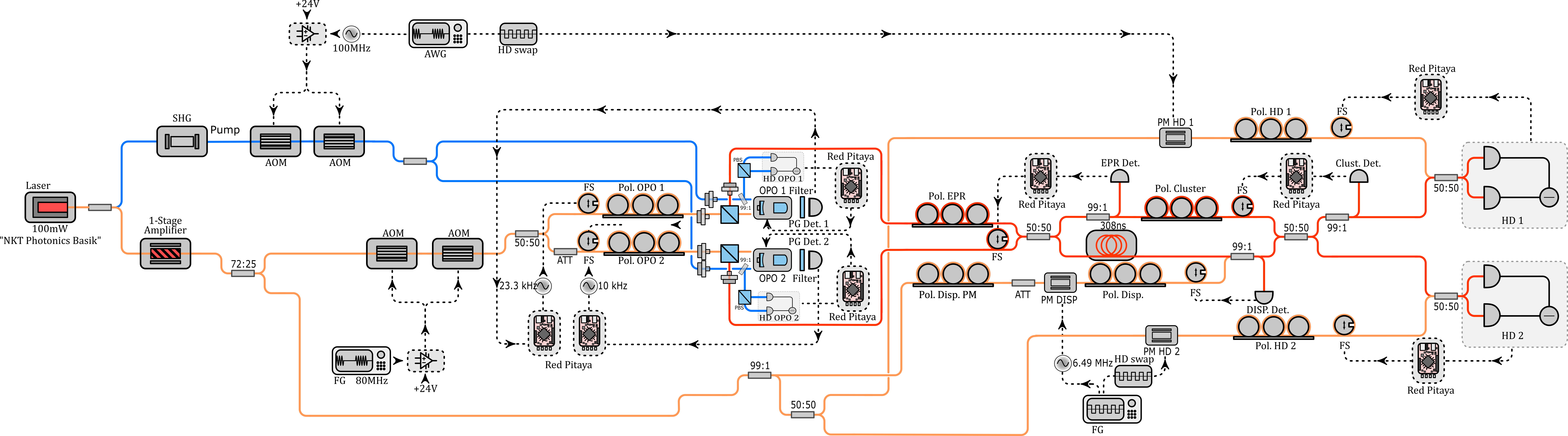}
  \caption{Experimental setup. Blue lines: 775\,nm pump; orange lines:
  1550\,nm probe, local oscillator (LO) and displacement beam; red lines:
  OPO outputs carrying the squeezed light; dashed black lines: electrical
  signals. SHG: second-harmonic-generation module; AOM: acousto-optic
  modulator; AWG: arbitrary waveform generator; FG: function generator;
  OPO: optical parametric oscillator; PBS: polarizing beam splitter;
  HD~OPO: homodyne detector for the OPO cavity lock; PG~Det.:
  parametric-gain detector for the pump--seed phase lock; FS: fiber
  stretcher; Pol.: fiber polarization controller; ATT: attenuator;
  PM: phase modulator; EPR~Det., Clust.~Det., DISP.~Det.: detectors for
  the EPR, cluster and displacement phase locks; HD: balanced homodyne
  detector. All locks are implemented digitally on Red Pitaya boards.}
  \label{fig:fullsetup}
\end{figure*}

The complete experimental setup is shown in Fig. \ref{fig:fullsetup}. All optical fields are derived from a single narrow-linewidth 1550 nm fiber laser (Koheras BASIK, NKT Photonics; Laser). Part of it is amplified in a single-stage fiber amplifier (1-Stage Amplifier) and used as the fundamental 1550 nm field and part is sent to a second-harmonic-generation module (SHG), which generates the 775 nm pump light. The 1550 nm light is divided into a probe path, which seeds the optical parametric oscillators (OPOs), and a local-oscillator (LO) path, which supplies the homodyne detectors and the displacement beam. Fiber polarization controllers (Pol.) set the polarization in each path.

The pump passes through two fiber-coupled acousto-optic modulators (AOM) in series, driven by a gated 100 MHz sine wave from an arbitrary waveform generator (AWG; Spectrum Instrumentation, model M4i.6631-x8) and amplified by high-power RF amplifiers. The second AOM reverses the frequency shift of the first, so the net frequency shift is zero. Gating the RF drive switches the pump on and off in a controlled sequence, which defines the 308 ns temporal modes in which squeezed light is generated. The pump is then split into two beams, which are coupled out to free space and directed to two hemilithic OPOs (OPO 1 and OPO 2), one beam each, where they drive spontaneous parametric down-conversion. Each OPO contains a periodically poled potassium titanyl phosphate (PPKTP) crystal and has a cavity bandwidth of $\approx$32 MHz. The pump power is $\approx$12 mW at OPO 1 and $\approx$14 mW at OPO 2, with thresholds of $\approx$20 mW and $\approx$30 mW, respectively.

The probe is gated by a second pair of AOMs, driven by a gated 80 MHz signal from a function generator (FG); again, the second AOM reverses the frequency shift of the first. This implements a sample-and-hold cycle that alternates between a 5 ms lock period and a 5 ms measurement period. During the lock period, the probe is on and all feedback loops stabilize the relative phases of the optical fields. During the measurement period, the probe is off, the controllers hold their last output, and data are acquired. All locks are digital and implemented on Red Pitaya FPGA controllers (Red Pitaya). After gating, the probe is split on a 50:50 coupler into two seed beams. Each seed passes through a home-built fiber stretcher (FS) based on a piezoelectric element, which imposes a phase modulation at $\approx$23.3 kHz (OPO 1) or $\approx$10 kHz (OPO 2). Fiber stretchers of the same type serve as the phase actuators in the locks described below. The seed of OPO 2 is attenuated by $\approx$95\% (ATT). One OPO operates in parametric amplification and the other in de-amplification, so this attenuation equalizes the probe powers emerging from the two OPOs.

Each OPO is stabilized by two locks. The cavity length is locked with a polarization-based DC scheme: a small fraction of the pump is tapped off before the OPO and overlapped with the pump light reflected from the cavity; the combined beam is split on a polarizing beam splitter (PBS) and measured with a homodyne detector (HD OPO 1, HD OPO 2), yielding an error signal that is fed back to the cavity length to keep the OPO resonant with the pump. The pump–seed phase is stabilized by a dither lock: the probe light transmitted through the cavity is measured with a parametric-gain detector (PG Det. 1, PG Det. 2), and the signal is demodulated at 23.3 kHz (OPO 1) or 10 kHz (OPO 2) and fed back to the fiber stretcher in the corresponding seed path. OPO 1 and OPO 2 are phase-locked to opposite extrema of the parametric gain, one to the maximum (amplification) and the other to the minimum (de-amplification). Their squeezed quadratures are therefore rotated by 90 degrees relative to each other.

The output of each OPO, containing both squeezed light and probe, is coupled back into fiber. The two outputs interfere on a 50:50 fiber coupler, generating a two-mode squeezed vacuum state, i.e. a finitely squeezed approximation of the Einstein–Podolsky–Rosen (EPR) state, which we refer to as the EPR state. One EPR mode is delayed by 308 ns, i.e. by one temporal mode, in $\approx$63 m of fiber. The delayed mode then interferes on a second 50:50 coupler with the undelayed mode of the subsequent EPR pair, so that consecutive pairs are entangled into a cluster state structure. The relative phase at each coupler is stabilized using 1\% of the light, which is tapped after the coupler with a 99:1 coupler and measured with a dedicated detector (EPR Det. for the first coupler, Clust. Det. for the second); the error signal is fed back to a fiber stretcher in one of the arms entering that coupler.

The displacement beam is derived from the LO path with a 99:1 coupler and attenuated by $\approx$99\% (ATT). It passes through a phase modulator (PM DISP) driven at 6.49 MHz by a function generator (FG). This places the displacement on the 6.49 MHz sideband, the same frequency at which the homodyne measurements are performed. The beam is injected into the delayed EPR mode after the delay line through a 99:1 coupler. Its phase is stabilized using the 23.3 kHz modulation, similarly to the locks above: the signal from the displacement-lock detector (DISP. Det.) is demodulated and fed back to a fiber stretcher in the displacement path.

The two outputs of the cluster-state coupler are measured by two balanced homodyne detectors (HD 1, HD 2). Each output is mixed with an LO on a 50:50 fiber coupler. The LOs are obtained by splitting the remaining LO light on a 50:50 coupler. Each LO passes through a phase modulator (PM HD 1, PM HD 2) and a fiber stretcher, which is used to lock the LO phase. The measured quadrature of each detector is set and switched by its phase modulator, driven by control waveforms (HD swap) from the AWG (for fast switching of the measured quadrature) and the FG (for slow switching). The detectors use photodiodes with $\approx$99\% quantum efficiency (Laser Components) and provide $\approx$16 dB of shot-noise clearance at 6.49 MHz with an LO power of $\approx$3.5 mW.

\subsection{Modes and measurement sequences}
\subsubsection{Temporal modes}
\label{sec: modes}

\begin{figure}[tbp]
  \centering
  \makebox[0.7\textwidth][l]{\textbf{(a)}}\\[-0.5ex]
  \includegraphics[width=0.7\textwidth]{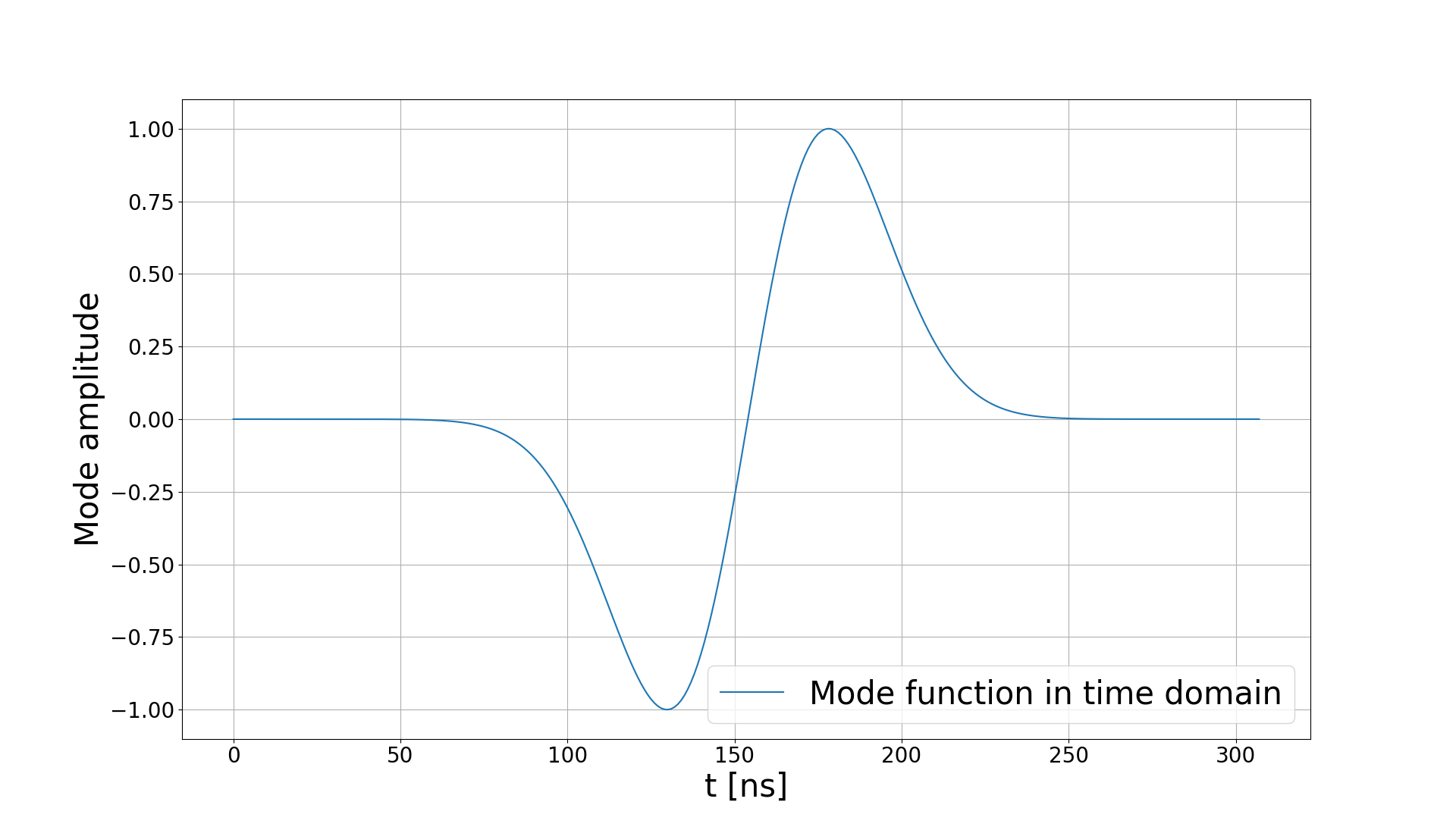}\\[0.5ex]
  \makebox[0.7\textwidth][l]{\textbf{(b)}}\\[-0.5ex]
  \includegraphics[width=0.7\textwidth]{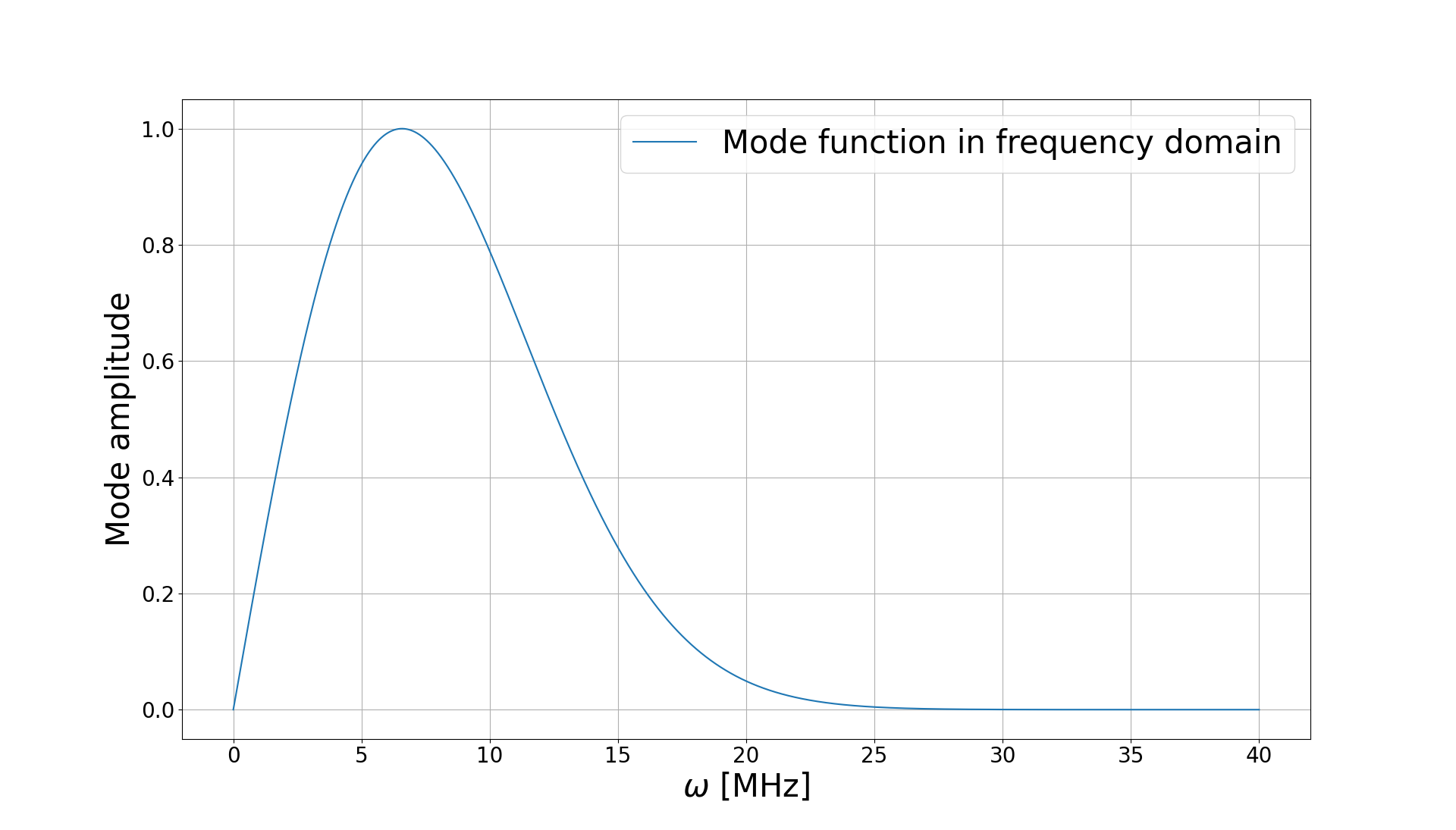}
  \caption{Mode function used to extract the quadrature values, (a) in the
  time domain and (b) in the frequency domain.}
  \label{fig:mode_function}
\end{figure}

The measurement frequency is chosen to be $f_\mathrm{m} \approx 6.49$~MHz.
This choice is dictated by the level of squeezing present around this
frequency and by the phase fluctuations that affect the measurement of the
cluster-state nullifiers at lower frequencies. To perform the measurement at
this frequency, the signal recorded with the oscilloscope (Teledyne LeCroy
WaveSurfer 3054) is multiplied in post-processing by a filter function (mode
function) and integrated over each temporal mode. The mode function is
\begin{equation}
  f(t) \propto \tau \cos(f_\mathrm{m}\tau)\,
  \exp\!\left[-\tfrac{1}{2}\,(2\pi f_\mathrm{m})^{2}\tau^{2}\right],
  \qquad \tau = t - t_{0},
  \label{eq:mode_function}
\end{equation}
for $|\tau| \le T_\mathrm{w}$ and zero otherwise, where $t_0$ is the centre of
the temporal mode. It is normalized to a peak amplitude of one and is shown in
the time domain in Fig.~\ref{fig:mode_function}a.

The length of the mode function is set by the length of the fiber delay, i.e.\ by the 308~ns duration of a temporal mode. The remaining parameters are chosen on the basis of the measured noise spectrum (Fig.~\ref{fig:mode_function}b), so that the unwanted frequency components are suppressed as far as possible.

\subsubsection{Measurements sequence}

The measurements follow the sample-and-hold cycle of a 5~ms lock period and a
5~ms measurement period. The first $\approx$2~ms of each measurement period
are discarded, because residual light from the probe beam, which serves as
the phase reference, is still present in the setup. The acquisition is then
triggered by the fast quadrature-switching waveform (HD swap) applied to the
phase modulator in the local-oscillator path of the homodyne detector. This
waveform is generated only during the measurement period and consists of a
repeated sequence of 20 temporal modes, corresponding to 6.16~$\mu$s. The
gating of the pump follows the same sequence, whereas the displacement
modulation is triggered only once, at the beginning of the measurement
period. Of the 6.16~$\mu$s of each sequence, 5~$\mu$s are recorded by the
oscilloscope. The delay between the trigger and the temporal modes in the
setup is carefully adjusted, so that the recorded traces are aligned with the
modes of the experiment. The optical delay of the temporal modes defined by the gated pump is likewise aligned with the modes defined by the switching of the homodyne measurement
quadrature. The data is acquired in the sequence mode of the
oscilloscope, in which one segment is recorded each time a trigger arrives,
until no trigger is detected for 100~$\mu$s. This yields 450 recorded
repetitions of the protocol in each acquisition, which is approximately the
minimum number required for the measured outcomes to follow a normal
distribution.

\subsection{Triggers and clock synchronization in the setup}
\label{sec: trig_clocks}

\subsubsection{Triggering system}
\label{sec: triggers}

All timing signals are derived from the function generator (FG) that drives the AOMs for the sample-and-hold gating of the probe light. One channel of this generator produces the AOM drive signal and the second channel produces the trigger signal. The trigger signal is split and distributed to three devices: the generator itself, where it serves as the external trigger of the gated AOM drive; the function generator that drives the displacement modulation and slowly switches the measurement quadrature of one homodyne detector (the detector that measures the $\hat{p}$ quadrature throughout the protocol); and the fast AWG that gates the pump and performs the fast switching of the measured quadrature. The trigger signal is generated as an arbitrary waveform, a square wave that is high for 5~ms and low for 5~ms, with the generator set to its ``TrueArb'' mode. This trigger configuration, together with the external clock described in Section~\ref{sec: clocks}, is essential: without it, the output channels of the function generators are not frequency-locked, which results in fast phase jitter and in a slow drift of the signal phase with respect to the trigger.

\subsubsection{Clock synchronization in the setup}
\label{sec: clocks}

A 10~MHz sinusoidal clock signal is generated by the arbitrary waveform generator built into the oscilloscope that is used to perform the measurements and acquire the data, and serves as the common reference for the other devices. The oscilloscope is used as the clock source because it has no clock input and therefore cannot be synchronized to an external reference. Without this synchronization, the frequency of the generated signals drifts away from the measurement frequency over time, and the signals from the function generators acquire a phase shift, so that the timing of the pulses in the experiment deviates from the defined measurement sequence.

\subsection{Analytical description of the protocol}
\label{sec:analytical_model}

In this section we will derive analytical expressions for the evolution of the state
through the experimental protocol, which are subsequently used to simulate the experiment. The derivation is divided into four steps. We first investigate the
ideal model without loss and phase noise, then include optical losses, then
phase noise, and finally combine both imperfections into the full model that is
compared with the experimental data.

\subsubsection{Ideal model - no loss and no phase noise}
\label{sec:analytical_model_ideal}

We begin with the ideal case, in which the protocol is free of loss and phase
noise and the evolution of the state is therefore unitary. At the beginning of protocol pump light, thus squeezing, is turned off and the input probe state at the initial moment (initial mode) of protocol is vacuum. Considering two input modes of experiment the covariance matrix and displacement vector read

\begin{equation}
V^{0}=\begin{pmatrix}1&0&0&0\\0&1&0&0\\0&0&1&0\\0&0&0&1\end{pmatrix},\qquad d_{0}=\begin{pmatrix}0\\0\\0\\0\end{pmatrix}.
\end{equation}

Two modes of this vacuum state interfere on beam splitter after which one of them undergoes the delay and gains displacement in P quadrature $\dd=(0,0,0,d_{p})^T$ to interfere with second vacuum mode again. After that Bell homodyne measurement is performed. Measured two modes are uncorrelated, so measuring the first one does not affect the second. The probe before any teleportation is
\begin{equation}
V^{IN}=\begin{pmatrix}1&0\\0&1\end{pmatrix},\qquad d_{IN}=\begin{pmatrix}0\\d_{p}\end{pmatrix} .
\label{eq:probe0}
\end{equation}

Next the pump squeezing is turned on and the two single mode squeezed vacuum (SMSV) states are generated.
\begin{equation}
V^{SMSV}=\begin{pmatrix}e^{-2r}&0&0&0\\0&e^{2r}&0&0\\0&0&e^{2r}&0\\0&0&0&e^{-2r}\end{pmatrix},\qquad d_{SMSV} \begin{pmatrix}0\\0\\0\\0\end{pmatrix}.
\label{eq:covariance_tmsv}
\end{equation}
Where $r$ is a squeezing parameter. Right after being generated they are combined on $50:50$ beam splitter to create two mode squeezed vacuum (TMSV), the finite-squeezing version of the EPR state:
\begin{equation}
V^{EPR}=\begin{pmatrix}\cosh 2r&0&\sinh 2r&0\\0&\cosh 2r&0&-\sinh 2r\\ \sinh 2r&0&\cosh 2r&0\\0&-\sinh 2r&0&\cosh 2r\end{pmatrix},\qquad
d_{EPR}=\begin{pmatrix}0\\0\\0\\0\end{pmatrix}.
\label{eq:VEPR}
\end{equation}

The input probe and the EPR (TMSV) pair are independent. Their joint state is
\begin{equation}
V^{IN}\oplus V^{EPR}=\begin{pmatrix}1&0&0&0&0&0\\0&1&0&0&0&0\\0&0&\cosh 2r&0&\sinh 2r&0\\0&0&0&\cosh 2r&0&-\sinh 2r\\
0&0&\sinh 2r&0&\cosh 2r&0\\0&0&0&-\sinh 2r&0&\cosh 2r\end{pmatrix},\qquad
d_{IN}\oplus d_{EPR} =\begin{pmatrix}0\\d_p\\0\\0\\0\\0\end{pmatrix}.
\end{equation}
One out of two modes, similarly to the initial probe, undergoes delay and displacement and second mode interfere with the input state (delayed and displaced mode of initial state) on the subsequent $50:50$ beam splitter. Its covariance matrix and displacement vector read

\begin{gather}
\wide{V^{CL}=\begin{pmatrix}
\cosh^2 r&0&-\sinh^2 r&0&\sqrt2\sinh r\cosh r&0\\
0&\cosh^2 r&0&-\sinh^2 r&0&-\sqrt2\sinh r\cosh r\\
-\sinh^2 r&0&\cosh^2 r&0&-\sqrt2\sinh r\cosh r&0\\
0&-\sinh^2 r&0&\cosh^2 r&0&\sqrt2\sinh r\cosh r\\
\sqrt2\sinh r\cosh r&0&-\sqrt2\sinh r\cosh r&0&\cosh 2r&0\\
0&-\sqrt2\sinh r\cosh r&0&\sqrt2\sinh r\cosh r&0&\cosh 2r
\end{pmatrix},}\nonumber\\
d_{CL}=\frac{1}{\sqrt2}\begin{pmatrix}0\\d_p\\0\\d_p\\0\\0\end{pmatrix}.
\label{eq:VCL}
\end{gather}

In the loss-less and phase noise free case nothing happens between the cluster beam splitter and the detectors, so the state going to this detector is
\begin{equation}
V^{HD}=V^{CL},\qquad d_{HD}=d_{CL} .
\end{equation}

A homodyne measurement on one mode of an entangled Gaussian state changes the
covariance matrix and the displacement vector of the non-measured mode. Writing
the state before the measurement as
\begin{equation}
\bm\sigma=\begin{pmatrix}\bm A&\bm C\\ \bm C^T&\bm B\end{pmatrix},\qquad
\bm\delta=\begin{pmatrix}\bm a\\ \bm b\end{pmatrix},
\label{eq:hdrule}
\end{equation}
where ($\bm A$, $\bm a$), ($\bm B$, $\bm b$) are the covariance matrix and the
displacement vector of the non-measured (measured) mode and $\bm C$ contains the
correlations between the two. The non-measured mode is left in the
state \cite{BraskGaussianToolbox}
\begin{align}
\bm\sigma' &= \bm A-\bm C\,(\bm\Pi\bm B\bm\Pi)^{-1}\bm C^T, \nonumber\\
\bm\delta' &= \bm a-\bm C\,(\bm\Pi\bm B\bm\Pi)^{-1}(\bm b-\bm u).
\end{align}
Here $u$ is the measurement outcome and
\begin{align}
\text{measurement of } \hat x: &\quad
\bm\Pi=\begin{pmatrix}1&0\\0&0\end{pmatrix}, \quad
\bm u=\begin{pmatrix}u\\0\end{pmatrix}, \quad
(\bm\Pi\bm B\bm\Pi)^{-1}=\frac{1}{B_{11}}\,\bm\Pi, \nonumber \\
\text{measurement of } \hat p: &\quad
\bm\Pi=\begin{pmatrix}0&0\\0&1\end{pmatrix}, \quad
\bm u=\begin{pmatrix}0\\u\end{pmatrix}, \quad
(\bm\Pi\bm B\bm\Pi)^{-1}=\frac{1}{B_{22}}\,\bm\Pi,
\label{eq:pimatrixhdrule}
\end{align}
where $B_{11}$ and $B_{22}$ are the diagonal entries of $\bm B$. Performing homodyne measurements on state $V^{CL}$ will modify it according to equations \ref{eq:hdrule} -- \ref{eq:pimatrixhdrule}. Performing Bell measurement is equivalent to modifying covariance matrix first with measurement of $X$ and then with measurement of $P$ (or symmetrically up to convention of beam splitters in reverse order).

For X measurement the measured mode is mode $2'$ (rows and columns 3, 4 of $\VHD$) and the non-measured modes are $1'$ and 3 (rows and columns 1, 2, 5, 6). The blocks of Eq.~\eqref{eq:hdrule} are
\begin{gather}
\bm A=\begin{pmatrix}\cosh^2 r&0&\sqrt2\sinh r\cosh r&0\\0&\cosh^2 r&0&-\sqrt2\sinh r\cosh r\\
\sqrt2\sinh r\cosh r&0&\cosh 2r&0\\0&-\sqrt2\sinh r\cosh r&0&\cosh 2r\end{pmatrix},\qquad
\bm B=\begin{pmatrix}\cosh^2 r&0\\0&\cosh^2 r\end{pmatrix},\nonumber\\
\bm C=\begin{pmatrix}-\sinh^2 r&0\\0&-\sinh^2 r\\-\sqrt2\sinh r\cosh r&0\\0&\sqrt2\sinh r\cosh r\end{pmatrix},\qquad
\bm a=\frac{1}{\sqrt2}\begin{pmatrix}0\\d_p\\0\\0\end{pmatrix},\qquad
\bm b=\frac{1}{\sqrt2}\begin{pmatrix}0\\d_p\end{pmatrix},\\
\bm C^T=\begin{pmatrix}-\sinh^2 r&0&-\sqrt2\sinh r\cosh r&0\\0&-\sinh^2 r&0&\sqrt2\sinh r\cosh r\end{pmatrix},\nonumber
\end{gather}
and, from Eq.~\eqref{eq:pimatrixhdrule} for a measurement of $\hat x$ with $B_{11}=\cosh^2 r$,
\begin{equation}
\bm\Pi=\begin{pmatrix}1&0\\0&0\end{pmatrix},\qquad \bm u=\begin{pmatrix}\xme{}\\0\end{pmatrix},\qquad
(\bm\Pi\bm B\bm\Pi)^{-1}=\frac{1}{\cosh^2 r}\,\bm\Pi .
\end{equation}
The two products from the equation \eqref{eq:hdrule} are
\begin{gather}
\bm C\,(\bm\Pi\bm B\bm\Pi)^{-1}\bm C^T=\frac{1}{\cosh^2 r}\begin{pmatrix}\sinh^4 r&0&\sqrt2\sinh^3 r\cosh r&0\\0&0&0&0\\
\sqrt2\sinh^3 r\cosh r&0&2\sinh^2 r\cosh^2 r&0\\0&0&0&0\end{pmatrix},\nonumber\\
\bm C\,(\bm\Pi\bm B\bm\Pi)^{-1}(\bm b-\bm u)=\frac{0-\xme{}}{\cosh^2 r}\begin{pmatrix}-\sinh^2 r\\0\\-\sqrt2\sinh r\cosh r\\0\end{pmatrix}
=\xme{}\begin{pmatrix}\tanh^2 r\\0\\ \sqrt2\tanh r\\0\end{pmatrix}.
\end{gather}
After X measurement the covariance matrix of $V^{CL'}$ state and displacement vector will read:
\begin{equation}
V^{CL'}=\begin{pmatrix}\dfrac{\cosh 2r}{\cosh^2 r}&0&\sqrt2\tanh r&0\\0&\cosh^2 r&0&-\sqrt2\sinh r\cosh r\\
\sqrt2\tanh r&0&1&0\\0&-\sqrt2\sinh r\cosh r&0&\cosh 2r\end{pmatrix},\qquad
d_{CL'}=\begin{pmatrix}-\tanh^2 r\; m_{x1}\\[2pt] \dfrac{d_p}{\sqrt2}\\[6pt] -\sqrt2\tanh r\;m_{x1}\\[2pt] 0\end{pmatrix}.
\label{eq:vclafterxideal}
\end{equation}

After which we apply change coming from P measurement, for which the measured mode is mode $1'$ (rows and columns 1, 2 of $V^{CL'}$ from equation \ref{eq:vclafterxideal}) and the non-measured mode is mode 3 (rows and columns 3, 4). Similarly to the previous case writing down all of needed elements present in equations \ref{eq:hdrule} -- \ref{eq:pimatrixhdrule}:
\begin{gather}
\bm A=\begin{pmatrix}1&0\\0&\cosh 2r\end{pmatrix},\qquad
\bm B=\begin{pmatrix}\dfrac{\cosh 2r}{\cosh^2 r}&0\\0&\cosh^2 r\end{pmatrix},\qquad
\bm C=\bm C^T=\begin{pmatrix}\sqrt2\tanh r&0\\0&-\sqrt2\sinh r\cosh r\end{pmatrix},\nonumber\\
\bm a=\begin{pmatrix}-\sqrt2\tanh r\;\xme{}\\0\end{pmatrix},\qquad
\bm b=\begin{pmatrix}-\tanh^2 r\;\xme{}\\[2pt] \dfrac{d_p}{\sqrt2}\end{pmatrix},
\end{gather}
and, from equation \ref{eq:pimatrixhdrule} for a measurement of $P$ with $B_{22}=\cosh^2 r$,
\begin{equation}
\bm\Pi=\begin{pmatrix}0&0\\0&1\end{pmatrix},\qquad \bm u=\begin{pmatrix}0\\ \pme{}\end{pmatrix},\qquad
(\bm\Pi\bm B\bm\Pi)^{-1}=\frac{1}{\cosh^2 r}\,\bm\Pi .
\end{equation}
The two products of the rule \eqref{eq:hdrule} are
\begin{gather}
\bm C\,(\bm\Pi\bm B\bm\Pi)^{-1}\bm C^T=\frac{1}{\cosh^2 r}\begin{pmatrix}0&0\\0&2\sinh^2 r\cosh^2 r\end{pmatrix}=\begin{pmatrix}0&0\\0&2\sinh^2 r\end{pmatrix},\nonumber\\
\bm C\,(\bm\Pi\bm B\bm\Pi)^{-1}(\bm b-\bm u)=\frac{1}{\cosh^2 r}\begin{pmatrix}0\\-\sqrt2\sinh r\cosh r\end{pmatrix}\left(\frac{d_p}{\sqrt2}-\pme{}\right)
=\begin{pmatrix}0\\-\tanh r\;d_p+\sqrt2\tanh r\;\pme{}\end{pmatrix}.
\end{gather}
After P measurement only mode 3 is left, with the new $V^{CL}$ and $d_{CL}$ being
\begin{equation}
V^{CL''}=\begin{pmatrix}1&0\\0&\cosh 2r-2\sinh^2 r\end{pmatrix}=\begin{pmatrix}1&0\\0&1\end{pmatrix},\qquad
d_{CL''}=\begin{pmatrix}-\sqrt2\tanh r\;m_{x1}\\ \tanh r\;d_p-\sqrt2\tanh r\;m_{p1}\end{pmatrix}.
\label{eq:ideal-afterP}
\end{equation}

The remaining mode passes the delay, which changes nothing, and is displaced in P by $d_p$. The covariance matrix and displacement vector of output state will be:
\begin{equation}
V^{OUT}=V^{CL''}=\begin{pmatrix}1&0\\0&1\end{pmatrix},\qquad
d_{OUT}=d_{CL''}+(0,d_{in})^{T}=\begin{pmatrix}-\sqrt2m_{x1}\tanh r\;\\ (1+\tanh r)\,d_p-\sqrt2 m_{p1}\tanh r\;\end{pmatrix}.
\label{eq:ideal-out}
\end{equation}

The output state is again a coherent state. Its displacement in P contains the new injected displacement $d_p$, the old one multiplied by $\tanh r$, and a displacement coming out of measuring out modes of entangled state. It is also an input state of the protocol for performing two teleportations. It is clear that teleportation does not change the covariance matrix of the input state as long as at the very beginning it is a vacuum state. Throughout the experimental protocol displacement only in one namely $P$ quadrature is object of interest. Thus, to make calculations easier any displacement in $X$ quadrature from now on will be set to zero.

Repeating all of the steps from above we can also derive the equations for displacements for higher number of teleportation
\begin{align}
d_{1}&=d_{IN}\tanh r-\sqrt2\tanh r\; m_{1}+\begin{pmatrix}0\\d_{p}\end{pmatrix}=\begin{pmatrix}0\\d_{p}\end{pmatrix}(1+\tanh r)-\sqrt2\tanh r\; m_{1},\\
d_{2}&=d_{1}\tanh r-\sqrt2\tanh r\; m_{2}+\begin{pmatrix}0\\d_{p}\end{pmatrix}=\begin{pmatrix}0\\d_{p}\end{pmatrix}(1+\tanh r+\tanh^2 r)-\sqrt2\tanh^2 r\; m_{1}-\sqrt2\tanh r\; m_{2},\\
d_{3}&=d_{2}\tanh r-\sqrt2\tanh r\; m_{3}+\begin{pmatrix}0\\d_{p}\end{pmatrix}\nonumber\\
&=\begin{pmatrix}0\\d_{p}\end{pmatrix}(1+\tanh r+\tanh^2 r+\tanh^3 r)-\sqrt2\tanh^3 r\; m_{1}-\sqrt2\tanh^2 r\; m_{2}-\sqrt2\tanh r\; m_{3},
\end{align}

Each teleportation introduces extra multiplication factor $\tanh r$ to displacement vector of an input state and adds a new term with the measurement dependent value and a new injected displacement
\begin{equation}
d_{M}=\begin{pmatrix}0\\d_{p}\end{pmatrix}\sum_{k=0}^{M}\tanh^{M-k}r-\sqrt2\sum_{n=1}^{M}\tanh^{M-n+1}r\; m_{n} .
\end{equation}
Using property of geometric sum $\sum_{k=0}^{M}q^k=(1-q^{M+1})/(1-q)$ we obtain expression for general displacement of the output state
\begin{equation}
\boxed{\;d_{M}=\begin{pmatrix}0\\d_{p}\end{pmatrix}\frac{1-\tanh^{M+1}r}{1-\tanh r}-\sqrt2\sum_{n=1}^{M}\tanh^{M-n+1}r\; m_{n},\qquad
m_{n}=\begin{pmatrix}m_{xn}\\ m_{pn}\end{pmatrix},\qquad
V_{M}=\begin{pmatrix}1&0\\0&1\end{pmatrix}.\;}
\label{eq:ideal-general}
\end{equation}

We finish the protocol with mixing the output state of last performed teleportation with non-delayed mode of freshly generated $EPR$ state followed by a readout of information which is implemented by performing homodyne measurements along the $P$ quadrature on both detectors. In order to write down evolution of the state we need to again apply the rule from equations \ref{eq:hdrule} -- \ref{eq:pimatrixhdrule}. Next we sum the outcomes of homodyne measurements and divide them by square root of two in order to obtain formula for the displacement that our input states undergoes
\begin{gather}
\hat d_{p}=\frac{1-\tanh r}{1-\tanh^{M+1}r}\left[z+\sqrt2\sum_{n=1}^{M}\tanh^{M-n+1}r\; m_{pn}\right],
\end{gather}
with
\begin{gather}
z\equiv\frac{m_{pfinal1}+m_{pfinal2}}{\sqrt2}
\end{gather}
where $m_{pfinal1},m_{pfinal2}$ are outcomes of final measurements of homodyne one and two respectively. We also can calculate standard deviation of displacement with respect to $z$
\begin{gather}
\sigma = \sqrt{Var(d_{p})}=\left(\frac{1-\tanh r}{1-\tanh^{M+1}r}\right)
\end{gather}
which represents uncertainty of measurements we perform or in other words sensitivity or precision that we can reach while performing measurement protocol.
\subsubsection{Model with losses included}
\label{sec:analytical_model_losses}

We now extend the ideal model from section \ref{sec:analytical_model_ideal} by considering linear optical losses. We include them in the model by adding fictitious beam splitter of transmission $\eta$ right before the homodyne detectors and vacuum in the unused port, followed by
tracing out the reflected light. It is equivalent to introducing equal losses both on the squeezing path as well as on the detected displacement path as opposed to considering their influence on the measurement protocol separately. In the initial state, the two SMSV states and the EPR pair are as in the lossless case: $V^{IN}$, $d_{IN}$ of equation \ref{eq:probe0} for the first teleportation, and $V^{EPR}$, $d_{EPR}$ of equation \ref{eq:VEPR}. We keep the convention of setting the displacement in $X$ quadrature to be zero. Thus the only new element is a fictitious beam splitter in front of each homodyne detector. Because of the loss the probe does not stay a coherent state, so the input state of a general teleportation is
\begin{equation}
V^{IN}=\begin{pmatrix}\nu&0\\0&\nu\end{pmatrix},\qquad d_{IN}=\begin{pmatrix}0\\d_{IN,p}\end{pmatrix},
\end{equation}
where $\nu$ is the variance of the probe (the same for X and P). For the first teleportation $\nu=1$ and $d_{IN}=\begin{pmatrix}0\\ d_{p}\end{pmatrix}$.

The input probe and the EPR pair are independent. Their joint state is
\begin{equation}
V^{IN}\oplus V^{EPR}=\begin{pmatrix}\nu&0&0&0&0&0\\0&\nu&0&0&0&0\\0&0&\cosh 2r&0&\sinh 2r&0\\0&0&0&\cosh 2r&0&-\sinh 2r\\
0&0&\sinh 2r&0&\cosh 2r&0\\0&0&0&-\sinh 2r&0&\cosh 2r\end{pmatrix},\qquad
d_{IN}\oplus d_{EPR}=\begin{pmatrix}0\\d_{IN,p}\\0\\0\\0\\0\end{pmatrix}.
\end{equation}

Input state and non-delayed mode of EPR state interfere on a beam splitter resulting in the following covariance matrix and displacement vector:
\begin{equation}
V^{CL}=\begin{pmatrix}
\frac{\nu+\cosh 2r}{2}&0&\frac{\nu-\cosh 2r}{2}&0&\frac{\sinh 2r}{\sqrt2}&0\\
0&\frac{\nu+\cosh 2r}{2}&0&\frac{\nu-\cosh 2r}{2}&0&-\frac{\sinh 2r}{\sqrt2}\\
\frac{\nu-\cosh 2r}{2}&0&\frac{\nu+\cosh 2r}{2}&0&-\frac{\sinh 2r}{\sqrt2}&0\\
0&\frac{\nu-\cosh 2r}{2}&0&\frac{\nu+\cosh 2r}{2}&0&\frac{\sinh 2r}{\sqrt2}\\
\frac{\sinh 2r}{\sqrt2}&0&-\frac{\sinh 2r}{\sqrt2}&0&\cosh 2r&0\\
0&-\frac{\sinh 2r}{\sqrt2}&0&\frac{\sinh 2r}{\sqrt2}&0&\cosh 2r
\end{pmatrix},\qquad
d_{CL}=\frac{1}{\sqrt2}\begin{pmatrix}0\\d_{IN,p}\\0\\d_{IN,p}\\0\\0\end{pmatrix}.
\end{equation}

For $\nu=1$ this is equal to the $V^{CL}$ of the ideal case from equation \ref{eq:VCL}.

Between the cluster beam splitter and the homodyne detectors the modes being measured are now experiencing loss when passing through fictitious beam splitter with transmission $\eta$. A fraction $\eta$ of the light is detected and $1-\eta$ fraction is replaced by vacuum. For a state with covariance matrix $V$ and displacement vector $d$ this behavior can be described by the loss operation
\begin{equation}
V\to L\,V\,L^T+N,\qquad d\to L\,d ,
\label{eq:lossop}
\end{equation}
where $L$ and $N$ are diagonal matrices. $L$ has $\sqrt\eta$ in entries related to modes that experience loss and $1$ in all other rows. $N$ has $1-\eta$ in entries related to modes that experience loss and $0$ in all other rows. Operator $L$ represents the partial loss of the signal and matrix $N$ is the vacuum noise that comes in with the loss. Here the modes that experience loss are modes measured by homodyne detectors (rows 1--4), and output mode (rows 5, 6) is not affected:
\begin{equation}
L=\begin{pmatrix}\sqrt\eta&0&0&0&0&0\\0&\sqrt\eta&0&0&0&0\\0&0&\sqrt\eta&0&0&0\\0&0&0&\sqrt\eta&0&0\\0&0&0&0&1&0\\0&0&0&0&0&1\end{pmatrix},\qquad
N=\begin{pmatrix}1-\eta&0&0&0&0&0\\0&1-\eta&0&0&0&0\\0&0&1-\eta&0&0&0\\0&0&0&1-\eta&0&0\\0&0&0&0&0&0\\0&0&0&0&0&0\end{pmatrix}.
\label{eq:LN}
\end{equation}
For the state going to the detectors we can write
\begin{equation}
V^{HD}=L\,V^{CL}L^T+N,\qquad d_{HD}=L\,d_{CL} .
\label{eq:loss-apply}
\end{equation}
Every entry of $V^{CL}$ gets a factor $\sqrt\eta$ for each detected mode it belongs to, and $1-\eta$ is added on the diagonal of the detected modes. With the variance of a detected quadrature

\begin{equation}
 \nu^{HD}=\frac{\eta}{2}\,(\nu+\cosh 2r)+1-\eta
\label{eq:loss-vardet}
\end{equation}
covariance matrix and displacement vector of the state at the input of the detectors read
\begin{gather}
V^{HD}=\begin{pmatrix}
 \nu^{HD}&0&\frac{\eta}{2}(\nu-\cosh 2r)&0&\sqrt{\frac{\eta}{2}}\sinh 2r&0\\
0& \nu^{HD}&0&\frac{\eta}{2}(\nu-\cosh 2r)&0&-\sqrt{\frac{\eta}{2}}\sinh 2r\\
\frac{\eta}{2}(\nu-\cosh 2r)&0& \nu^{HD}&0&-\sqrt{\frac{\eta}{2}}\sinh 2r&0\\
0&\frac{\eta}{2}(\nu-\cosh 2r)&0& \nu^{HD}&0&\sqrt{\frac{\eta}{2}}\sinh 2r\\
\sqrt{\frac{\eta}{2}}\sinh 2r&0&-\sqrt{\frac{\eta}{2}}\sinh 2r&0&\cosh 2r&0\\
0&-\sqrt{\frac{\eta}{2}}\sinh 2r&0&\sqrt{\frac{\eta}{2}}\sinh 2r&0&\cosh 2r
\end{pmatrix},\nonumber\\
d_{HD}=\sqrt{\frac{\eta}{2}}\begin{pmatrix}0\\d_{IN,p}\\0\\d_{IN,p}\\0\\0\end{pmatrix}.
\label{eq:loss-VHD}
\end{gather}
The output mode does not experience loss, so its block stays $\cosh 2r$. The loss reaches the output mode only through the measurement.

Similarly to lossless case we perform now Bell measurement for which we first perform measurement along the $X$ quadrature followed by measurement of $P$ quadrature. Applying the rule from equations \ref{eq:hdrule} -- \ref{eq:pimatrixhdrule} for the X measurement (on mode representad by rows and columns 3,4 of the $V^{HD}$) we obtain for the two non-measured modes
\begin{gather}
V^{HD'}=\begin{pmatrix}
\dfrac{(\eta\cosh 2r+1-\eta)(\eta\nu+1-\eta)}{ \nu^{HD}}&0&\sqrt{\dfrac{\eta}{2}}\sinh 2r\,\dfrac{\eta\nu+1-\eta}{ \nu^{HD}}&0\\[8pt]
0& \nu^{HD}&0&-\sqrt{\dfrac{\eta}{2}}\sinh 2r\\[8pt]
\sqrt{\dfrac{\eta}{2}}\sinh 2r\,\dfrac{\eta\nu+1-\eta}{ \nu^{HD}}&0&\cosh 2r-\dfrac{\eta\sinh^2 2r}{2 \nu^{HD}}&0\\[8pt]
0&-\sqrt{\dfrac{\eta}{2}}\sinh 2r&0&\cosh 2r
\end{pmatrix},\nonumber\\
d_{HD'}=\begin{pmatrix}\dfrac{\eta\,(\nu-\cosh 2r)}{2 \nu^{HD}}\;m_{x1}\\[8pt]
\sqrt{\dfrac{\eta}{2}}\,d_{IN,p}\\[8pt]
-\sqrt{\dfrac{\eta}{2}}\,\dfrac{\sinh 2r}{ \nu^{HD}}\;m_{x1}\\[8pt]
0\end{pmatrix} = \begin{pmatrix}0\\[8pt]
\sqrt{\dfrac{\eta}{2}}\,d_{IN,p}\\[8pt]
0\\[8pt]
0\end{pmatrix}.
\label{eq:loss-afterX}
\end{gather}
Applying the same rule for the P measurement (rows and columns 1, 2 of updated $V^{HD'}$ from equation \ref{eq:loss-afterX}), we obtain the expression describing the output mode
\begin{equation}
V^{HD''}=\left(\cosh 2r-\frac{\eta\sinh^2 2r}{2 \nu^{HD}}\right)\begin{pmatrix}1&0\\0&1\end{pmatrix},\qquad
d_{HD''}=\begin{pmatrix}0\\[8pt]
d_{IN,p}\,\dfrac{\eta\sinh 2r}{2 \nu^{HD}}-\sqrt{\dfrac{\eta}{2}}\,\dfrac{\sinh 2r}{ \nu^{HD}}\;m_{p1}\end{pmatrix}.
\label{eq:loss-afterP}
\end{equation}
The same factor is present both in the covariance matrix and in the displacement vector. We define it as the gain of single teleportation,
\begin{equation}
g=\frac{\eta\sinh 2r}{2 \nu^{HD}}=\frac{\sinh 2r}{\nu+\cosh 2r+2\frac{1-\eta}{\eta}} ,
\label{eq:loss-g}
\end{equation}
where the second form follows from equation \ref{eq:loss-vardet} after dividing numerator and denominator by $\eta$. The output mode is displaced in P by extra $d_{p}$, analogically to previous case. The covariance matrix and displacement vector of the output state are
\begin{equation}
V^{OUT}=(\cosh 2r-g\sinh 2r\,)\begin{pmatrix}1&0\\0&1\end{pmatrix},\qquad
d_{OUT}=\begin{pmatrix}0\\[8pt]
d_{IN,p}\,g-\sqrt{\dfrac{2}{\eta}}\;g\;m_{p1}+d_{p}\end{pmatrix}.
\label{eq:loss-map}
\end{equation}
The variance of the probe changes from $\nu$ to $\cosh 2r-\sinh 2r\,g$. The P entry of $d_{OUT}$ consists of term proportional to the displacement of the input state, displacement coming out of measuring cluster modes and and the newly injected displacement. The X entry is set to zero, in line with the convention of zero displacement in $X$. It is easy to check that for $\eta=1$ and $\nu=1$ the gain is $g=\sinh 2r/(1+\cosh 2r)=\tanh r$ and $V^{OUT}$ is the identity matrix, as in the lossless case.

For the first teleportation the input state is the probe from the equation \ref{eq:probe0}, for which $\nu=1$ and $d_{IN,p}=d_{p}$. Knowing that we can calcualte $ \nu^{HD}=1+\eta\sinh^2 r$ and write the gain form equation \ref{eq:loss-g} as
\begin{equation}
g_1=\frac{\eta\,\sinh r\cosh r}{1+\eta\,\sinh^2 r} .
\label{eq:g1}
\end{equation}

Following the experimental protocol the output state of one teleportation is the input state of the next one. We set variance of the output state after $n$ teleportation to $\nu_n$, its displacement to be the $d_{n,p}$, and $g_n$ the gain related to teleportation $n$. The probe before any teleportation (equation \ref{eq:probe0}) has $\nu_0=1$ and $d_{0,p}=d_{p}$. From equations \ref{eq:loss-g} and \ref{eq:loss-map} we write the general formula for calculating gain of teleportation and variance and displacement of output state after teleportation:
\begin{gather}
g_n=\frac{\sinh 2r}{\nu_{n-1}+\cosh 2r+2\frac{1-\eta}{\eta}},\qquad \nu_n=\cosh 2r-\sinh 2r\;g_n,\nonumber\\
d_{n,p}=d_{n-1,p}\,g_n-\sqrt{\frac{2}{\eta}}\;g_n\; m_{pn}+d_{p} .
\label{eq:loss-rec}
\end{gather}
For one, two and three teleportations this gives
\begin{align}
d_{1,p}&=d_{p}(1+g_1)-\sqrt{\frac{2}{\eta}}\;g_1\; m_{p1},\nonumber\\
d_{2,p}&=d_{p}(1+g_2+g_2g_1)-\sqrt{\frac{2}{\eta}}\left(g_2g_1\; m_{p1}+g_2\; m_{p2}\right),\nonumber\\
d_{3,p}&=d_{p}(1+g_3+g_3g_2+g_3g_2g_1)-\sqrt{\frac{2}{\eta}}\left(g_3g_2g_1\; m_{p1}+g_3g_2\; m_{p2}+g_3\; m_{p3}\right),
\end{align}
and for $M$ teleportations
\begin{equation}
d_{M,p}=d_{p}\Big(1+g_M+g_Mg_{M-1}+\dots+g_M\cdots g_1\Big)-\sqrt{\frac{2}{\eta}}\sum_{n=1}^{M}\big(g_M\cdots g_{n}\big)\; m_{pn} .
\label{eq:loss-unrolled}
\end{equation}

Defining $y_n=\nu_n+\cosh 2r+2\frac{1-\eta}{\eta}$ equation \ref{eq:loss-rec} reads
\begin{equation}
g_n=\frac{\sinh 2r}{y_{n-1}},\qquad
y_n=2\cosh 2r+2\,\frac{1-\eta}{\eta}-\frac{\sinh^2 2r}{y_{n-1}},\qquad y_0=1+\cosh 2r+2\,\frac{1-\eta}{\eta} .
\end{equation}
The recursion can be solved to make it easier to simulate and understand the behavior of the system. We put
\begin{equation}
y_n=\sinh 2r\,\frac{D_{n+1}}{D_n},\qquad
y_{n-1}=\sinh 2r\,\frac{D_n}{D_{n-1}},
\end{equation}
so that the expression for the gain becomes
\begin{equation}
g_n=\frac{\sinh 2r}{y_{n-1}}=\frac{D_{n-1}}{D_n}.
\label{eq:loss-gD}
\end{equation}
Followed by inserting $y_n$ and $y_{n-1}$ into the recursion gives
\begin{equation}
\sinh 2r\,\frac{D_{n+1}}{D_n}=2\cosh 2r+2\,\frac{1-\eta}{\eta}-\sinh 2r\,\frac{D_{n-1}}{D_n},
\end{equation}
and next multiplying by $D_n/\sinh 2r$ we derive
\begin{equation}
D_{n+1}=2\kappa\,D_n-D_{n-1},\qquad \kappa=\frac{\cosh 2r+\frac{1-\eta}{\eta}}{\sinh 2r}.
\label{eq:loss-kappa}
\end{equation}
We look for solution in the form of a power:
\begin{equation}
D_n=\lambda^{n}\qquad\Rightarrow\qquad \lambda^2-2\kappa\lambda+1=0 .
\end{equation}
The two solutions for this equation are $\lambda$ and $1/\lambda$, with
\begin{equation}
\lambda=\kappa-\sqrt{\kappa^2-1},\qquad \frac{1}{\lambda}=\kappa+\sqrt{\kappa^2-1},\qquad \lambda+\frac{1}{\lambda}=2\kappa .
\label{eq:loss-lambda}
\end{equation}
In the lossless case $\lambda$ is equal to $\tanh r$. In the presence of loss it is smaller:
\begin{equation}
\eta=1:\quad \kappa=\frac{\cosh 2r}{\sinh 2r},\quad \lambda=\frac{\cosh 2r-1}{\sinh 2r}=\tanh r ,
\qquad\qquad \eta<1:\quad 0<\lambda<\tanh r .
\end{equation}
The general solution of equation \ref{eq:loss-kappa} reads
\begin{equation}
D_n=A\,\lambda^{-n}+B\,\lambda^{n} ,
\end{equation}
in which the constants $A$ and $B$ can be calculated from $y_0$:
\begin{equation}
\frac{D_1}{D_0}=\frac{y_0}{\sinh 2r}=\frac{1+\cosh 2r+2\,\frac{1-\eta}{\eta}}{\sinh 2r}=2\kappa-\frac{\cosh 2r-1}{\sinh 2r}=2\kappa-\tanh r .
\end{equation}
Inserting $D_0=A+B$, $D_1=A/\lambda+B\lambda$ and $2\kappa=\lambda+1/\lambda$ we arrive at:
\begin{gather}
\frac{A}{\lambda}+B\lambda=\left(\lambda+\frac{1}{\lambda}-\tanh r\right)(A+B),\nonumber\\
A\,(\tanh r-\lambda)=B\left(\frac{1}{\lambda}-\tanh r\right),\nonumber\\
B=A\,\lambda\,\epsilon,\qquad \epsilon=\frac{\tanh r-\lambda}{1-\lambda\tanh r} .
\label{eq:loss-eps}
\end{gather}
We then realize that only ratios of $D_n$ are needed, so without the loss of generality we can set $A=1$:
\begin{equation}
D_n=\lambda^{-n}+\epsilon\,\lambda^{n+1} .
\label{eq:loss-Dn}
\end{equation}
The gain and the variance of the probe after teleportation $n$ become
\begin{gather}
g_n=\frac{D_{n-1}}{D_n}=\frac{\lambda^{-(n-1)}+\epsilon\,\lambda^{n}}{\lambda^{-n}+\epsilon\,\lambda^{n+1}}
=\lambda\,\frac{1+\epsilon\,\lambda^{2n-1}}{1+\epsilon\,\lambda^{2n+1}} .
\label{eq:loss-gn}
\end{gather}
And the product of the gain factors from different teleportation equation \ref{eq:loss-unrolled} is
\begin{align}
g_Mg_{M-1}\cdots g_{n}&=\frac{D_{M-1}}{D_M}\cdot\frac{D_{M-2}}{D_{M-1}}\cdots\frac{D_{n-1}}{D_n}=\frac{D_{n-1}}{D_M}\nonumber\\
&=\lambda^{M-n+1}\,\frac{1+\epsilon\,\lambda^{2n-1}}{1+\epsilon\,\lambda^{2M+1}} ,
\label{eq:loss-prod}
\end{align}
Which let us calculate sum of products of gain coefficients in equation \ref{eq:loss-unrolled}
\begin{align}
G_M&=1+g_M+g_Mg_{M-1}+\dots+g_M\cdots g_1=\frac{1}{D_M}\sum_{k=0}^{M}D_k\nonumber\\
&=\frac{1}{D_M}\left[\sum_{k=0}^{M}\lambda^{-k}+\epsilon\,\lambda\sum_{k=0}^{M}\lambda^{k}\right]
=\frac{1}{D_M}\cdot\frac{1-\lambda^{M+1}}{1-\lambda}\left[\lambda^{-M}+\epsilon\,\lambda\right]\nonumber\\
&=\frac{1-\lambda^{M+1}}{1-\lambda}\cdot\frac{1+\epsilon\,\lambda^{M+1}}{1+\epsilon\,\lambda^{2M+1}} .
\label{eq:loss-GM}
\end{align}
The displacement $d_{M,p}$ and the variance $\nu_M$ of the probe after $M$ teleportations are therefore
\begin{gather}
d_{M,p}=d_{p}\,G_M-\sqrt{\frac{2}{\eta}}\sum_{n=1}^{M}\lambda^{M-n+1}\,\frac{1+\epsilon\,\lambda^{2n-1}}{1+\epsilon\,\lambda^{2M+1}}\; m_{pn},\nonumber\\
\nu_M=\cosh 2r-\lambda\sinh 2r\,\frac{1+\epsilon\,\lambda^{2M-1}}{1+\epsilon\,\lambda^{2M+1}} .
\label{eq:loss-general}
\end{gather}

Following the same pattern as in the ideal case for the readout of information, we derive the equation for the output of the final measurements, however this time including losses:
\begin{equation}
z\equiv\frac{m_{pfinal1}+m_{pfinal2}}{\sqrt{2\eta}},\qquad \langle z\rangle=d_{M,p},\qquad \mathrm{Var}(z)=\nu_M+\frac{1-\eta}{\eta} ,
\label{eq:loss-readout}
\end{equation}
where $m_{pfinal1}$, $m_{pfinal2}$ are outcomes of final measurements of homodyne one and two respectively, and $\frac{1-\eta}{\eta}$ is the extra noise from the loss in front of the detectors. Inserting $d_{M,p}$ from equation \ref{eq:loss-general} we obtain the formula for the displacement that our input state undergoes and its standard deviation. The complete result of the final measurement, together with all variables it contains, is
\begin{equation}
\setlength{\fboxsep}{10pt}
\boxed{\begin{gathered}
\hat d_{p}=\frac{1}{G_M}\left[z+\sqrt{\frac{2}{\eta}}\sum_{n=1}^{M}\lambda^{M-n+1}\,\frac{1+\epsilon\,\lambda^{2n-1}}{1+\epsilon\,\lambda^{2M+1}}\; m_{pn}\right],\\[4pt]
\sigma=\sqrt{\mathrm{Var}(\hat d_{p})}=\frac{1}{G_M}\sqrt{\nu_M+\frac{1-\eta}{\eta}},\\[10pt]
z=\frac{m_{pfinal1}+m_{pfinal2}}{\sqrt{2\eta}},\\[4pt]
G_M=\frac{1-\lambda^{M+1}}{1-\lambda}\cdot\frac{1+\epsilon\,\lambda^{M+1}}{1+\epsilon\,\lambda^{2M+1}},\\[4pt]
\nu_M=\cosh 2r-\lambda\sinh 2r\,\frac{1+\epsilon\,\lambda^{2M-1}}{1+\epsilon\,\lambda^{2M+1}},\\[4pt]
\lambda=\kappa-\sqrt{\kappa^2-1},\qquad
\kappa=\frac{\cosh 2r+\frac{1-\eta}{\eta}}{\sinh 2r},\qquad
\epsilon=\frac{\tanh r-\lambda}{1-\lambda\tanh r}
\end{gathered}}
\label{eq:loss-estimate}
\end{equation}
Here $\hat d_{p}$ is the estimate of the injected displacement and $\sigma$ is its standard deviation, $M$ is the number of teleportations, $m_{pn}$ is the outcome of the homodyne measurement of the $P$ quadrature in teleportation $n$, $m_{pfinal1}$ and $m_{pfinal2}$ are the outcomes of the final measurements, $r$ is the squeezing parameter and $\eta$ is the transmission of the fictitious beam splitter in front of each homodyne detector.

For the ideal case, $\eta=1$, we have
\begin{equation}
\lambda=\tanh r,\qquad \epsilon=0,\qquad \nu_M=1,\qquad G_M=\frac{1-\tanh^{M+1}r}{1-\tanh r},
\end{equation}
and the expressions above become those of section \ref{sec:analytical_model_ideal}:
\begin{equation}
\hat d_{p}=\frac{1-\tanh r}{1-\tanh^{M+1}r}\left[z+\sqrt2\sum_{n=1}^{M}\tanh^{M-n+1}r\; m_{pn}\right],\qquad
\sigma=\frac{1-\tanh r}{1-\tanh^{M+1}r} .
\label{eq:loss-ideal-limit}
\end{equation}

\subsubsection{Model with phase noise included}
\label{sec:analytical_model_phase_noise}

In this section we come back to the lossless case from section \ref{sec:analytical_model_ideal} and derive model that includes phase noise. We model the phase noise as a phase shift $\phi$ which is the same at four places of the experiment namely: on one of the two SMSV states they interfere and create EPR state, on one of modes of EPR state the cluster beam splitter, and on both measured modes right before the homodyne detectors. A phase shift rotates the quadratures of a mode. For a state with covariance matrix $V$ and displacement vector $d$ this behavior can be described by the rotation operation
\begin{equation}
V\to R\,V\,R^T,\qquad d\to R\,d ,\qquad R=\begin{pmatrix}\cos\phi&\sin\phi\\-\sin\phi&\cos\phi\end{pmatrix}.
\label{eq:phaseop}
\end{equation}
The rotation mixes the $X$ and $P$ quadratures. In contrast to the previous cases the displacement in $X$ quadrature can therefore not be set to zero, and the probe does not stay a coherent state. The input state of a general teleportation is
\begin{equation}
V^{IN}=\begin{pmatrix}V^{IN}_{xx}&V^{IN}_{xp}\\ V^{IN}_{xp}&V^{IN}_{pp}\end{pmatrix},\qquad d_{IN}=\begin{pmatrix}d_{IN,x}\\d_{IN,p}\end{pmatrix}.
\label{eq:phase-input}
\end{equation}
For the first teleportation it is the probe from equation \ref{eq:probe0}, $V^{IN}=\begin{pmatrix}1&0\\0&1\end{pmatrix}$ and $d_{IN}=\begin{pmatrix}0\\ d_{p}\end{pmatrix}$.

The two SMSV states are generated as in equation \ref{eq:covariance_tmsv}. The strength of the correlations of the EPR state depends on the relative phase between the two squeezed states. After this rotation the covariance matrix and displacement vector of the two SMSV states read
\begin{gather}
V^{SMSV'}=\begin{pmatrix}
e^{-2r}&0&0&0\\
0&e^{2r}&0&0\\
0&0&e^{2r}\cos^2\phi+e^{-2r}\sin^2\phi&-\left(e^{2r}-e^{-2r}\right)\sin\phi\cos\phi\\
0&0&-\left(e^{2r}-e^{-2r}\right)\sin\phi\cos\phi&e^{2r}\sin^2\phi+e^{-2r}\cos^2\phi
\end{pmatrix},\nonumber\\
d_{SMSV'}=\begin{pmatrix}0\\0\\0\\0\end{pmatrix}.
\label{eq:phase-SMSV}
\end{gather}
The two SMSV states are then combined on the beam splitter. The covariance matrix and displacement vector of the EPR state read
\begin{gather}
V^{EPR}=\begin{pmatrix}
\cosh 2r-\sinh 2r\sin^2\phi&-\sinh 2r\sin\phi\cos\phi&\sinh 2r\cos^2\phi&-\sinh 2r\sin\phi\cos\phi\\
-\sinh 2r\sin\phi\cos\phi&\cosh 2r+\sinh 2r\sin^2\phi&-\sinh 2r\sin\phi\cos\phi&-\sinh 2r\cos^2\phi\\
\sinh 2r\cos^2\phi&-\sinh 2r\sin\phi\cos\phi&\cosh 2r-\sinh 2r\sin^2\phi&-\sinh 2r\sin\phi\cos\phi\\
-\sinh 2r\sin\phi\cos\phi&-\sinh 2r\cos^2\phi&-\sinh 2r\sin\phi\cos\phi&\cosh 2r+\sinh 2r\sin^2\phi
\end{pmatrix},\nonumber\\
d_{EPR}=\begin{pmatrix}0\\0\\0\\0\end{pmatrix}.
\label{eq:phase-VEPR}
\end{gather}
For $\phi=0$ this is equal to the $V^{EPR}$ of the ideal case from equation \ref{eq:VEPR}.

The input state from equation \ref{eq:phase-input} is rotated by $\phi$ before the cluster beam splitter. After this rotation its covariance matrix and displacement vector read
\begin{gather}
W=\begin{pmatrix}
V^{IN}_{xx}\cos^2\phi+2V^{IN}_{xp}\sin\phi\cos\phi+V^{IN}_{pp}\sin^2\phi&\left(V^{IN}_{pp}-V^{IN}_{xx}\right)\sin\phi\cos\phi+V^{IN}_{xp}\left(\cos^2\phi-\sin^2\phi\right)\\[4pt]
\left(V^{IN}_{pp}-V^{IN}_{xx}\right)\sin\phi\cos\phi+V^{IN}_{xp}\left(\cos^2\phi-\sin^2\phi\right)&V^{IN}_{xx}\sin^2\phi-2V^{IN}_{xp}\sin\phi\cos\phi+V^{IN}_{pp}\cos^2\phi
\end{pmatrix},\nonumber\\
w=\begin{pmatrix}d_{IN,x}\cos\phi+d_{IN,p}\sin\phi\\-d_{IN,x}\sin\phi+d_{IN,p}\cos\phi\end{pmatrix}.
\label{eq:phase-W}
\end{gather}
For the first teleportation
\begin{equation}
V^{IN}=\begin{pmatrix}1&0\\0&1\end{pmatrix},\qquad d_{IN}=\begin{pmatrix}0\\ d_{p}\end{pmatrix},
\end{equation}
which gives
\begin{equation}
W=\begin{pmatrix}1&0\\0&1\end{pmatrix},\qquad w=\begin{pmatrix}d_{p}\sin\phi\\ d_{p}\cos\phi\end{pmatrix}.
\label{eq:phase-W1}
\end{equation}
To make calculations easier to follow, we denote the four different entries of $V^{EPR}$ by
\begin{gather}
e_x=\cosh 2r-\sinh 2r\sin^2\phi,\qquad e_p=\cosh 2r+\sinh 2r\sin^2\phi,\nonumber\\
e_c=-\sinh 2r\sin\phi\cos\phi,\qquad k=\sinh 2r\cos^2\phi ,
\label{eq:phase-e}
\end{gather}
and the entries of covariance matrix $W$ and displacement vector $w$ by
\begin{equation}
W=\begin{pmatrix}W_{xx}&W_{xp}\\W_{xp}&W_{pp}\end{pmatrix},\qquad w=\begin{pmatrix}w_x\\w_p\end{pmatrix}.
\label{eq:phase-W-short}
\end{equation}
The rotated input state and the EPR pair are independent. Their joint state is
\begin{equation}
W\oplus V^{EPR}=\begin{pmatrix}W_{xx}&W_{xp}&0&0&0&0\\W_{xp}&W_{pp}&0&0&0&0\\0&0&e_x&e_c&k&e_c\\0&0&e_c&e_p&e_c&-k\\
0&0&k&e_c&e_x&e_c\\0&0&e_c&-k&e_c&e_p\end{pmatrix},\qquad
w\oplus d_{EPR}=\begin{pmatrix}w_x\\w_p\\0\\0\\0\\0\end{pmatrix}.
\end{equation}
The input state and non-delayed mode of EPR state interfere on a beam splitter resulting in the following:
\begin{equation}
V^{CL}=\begin{pmatrix}
\frac{W_{xx}+e_x}{2}&\frac{W_{xp}+e_c}{2}&\frac{W_{xx}-e_x}{2}&\frac{W_{xp}-e_c}{2}&\frac{k}{\sqrt2}&\frac{e_c}{\sqrt2}\\[2pt]
\frac{W_{xp}+e_c}{2}&\frac{W_{pp}+e_p}{2}&\frac{W_{xp}-e_c}{2}&\frac{W_{pp}-e_p}{2}&\frac{e_c}{\sqrt2}&-\frac{k}{\sqrt2}\\[2pt]
\frac{W_{xx}-e_x}{2}&\frac{W_{xp}-e_c}{2}&\frac{W_{xx}+e_x}{2}&\frac{W_{xp}+e_c}{2}&-\frac{k}{\sqrt2}&-\frac{e_c}{\sqrt2}\\[2pt]
\frac{W_{xp}-e_c}{2}&\frac{W_{pp}-e_p}{2}&\frac{W_{xp}+e_c}{2}&\frac{W_{pp}+e_p}{2}&-\frac{e_c}{\sqrt2}&\frac{k}{\sqrt2}\\[2pt]
\frac{k}{\sqrt2}&\frac{e_c}{\sqrt2}&-\frac{k}{\sqrt2}&-\frac{e_c}{\sqrt2}&e_x&e_c\\[2pt]
\frac{e_c}{\sqrt2}&-\frac{k}{\sqrt2}&-\frac{e_c}{\sqrt2}&\frac{k}{\sqrt2}&e_c&e_p
\end{pmatrix},\qquad
d_{CL}=\frac{1}{\sqrt2}\begin{pmatrix}w_x\\w_p\\w_x\\w_p\\0\\0\end{pmatrix}.
\label{eq:phase-VCL}
\end{equation}
Between the cluster beam splitter and the homodyne detectors we apply once more a rotation by $\phi$ to the modes that are being measured. For the state at ten input of the detectors we can write
\begin{equation}
V^{HD}=R^{HD}\,V^{CL}\left(R^{HD}\right)^T,\qquad d_{HD}=R^{HD}\,d_{CL} ,\qquad
R^{HD}=\begin{pmatrix}\cos\phi&\sin\phi&0&0&0&0\\-\sin\phi&\cos\phi&0&0&0&0\\0&0&\cos\phi&\sin\phi&0&0\\0&0&-\sin\phi&\cos\phi&0&0\\0&0&0&0&1&0\\0&0&0&0&0&1\end{pmatrix}.
\label{eq:phase-apply}
\end{equation}

Similarly to the previous cases we perform now Bell measurement. Firstly we do the X measurement on mode $2'$ (rows and columns 3, 4 of $V^{HD}$) and then the P measurement on mode $1'$ (rows and columns 1, 2). The covariance matrix and displacement vector before the homodyne measurements read
\begin{gather}
V^{HD}=R^{HD}\begin{pmatrix}
\frac{W_{xx}+e_x}{2}&\frac{W_{xp}+e_c}{2}&\frac{W_{xx}-e_x}{2}&\frac{W_{xp}-e_c}{2}&\frac{k}{\sqrt2}&\frac{e_c}{\sqrt2}\\[2pt]
\frac{W_{xp}+e_c}{2}&\frac{W_{pp}+e_p}{2}&\frac{W_{xp}-e_c}{2}&\frac{W_{pp}-e_p}{2}&\frac{e_c}{\sqrt2}&-\frac{k}{\sqrt2}\\[2pt]
\frac{W_{xx}-e_x}{2}&\frac{W_{xp}-e_c}{2}&\frac{W_{xx}+e_x}{2}&\frac{W_{xp}+e_c}{2}&-\frac{k}{\sqrt2}&-\frac{e_c}{\sqrt2}\\[2pt]
\frac{W_{xp}-e_c}{2}&\frac{W_{pp}-e_p}{2}&\frac{W_{xp}+e_c}{2}&\frac{W_{pp}+e_p}{2}&-\frac{e_c}{\sqrt2}&\frac{k}{\sqrt2}\\[2pt]
\frac{k}{\sqrt2}&\frac{e_c}{\sqrt2}&-\frac{k}{\sqrt2}&-\frac{e_c}{\sqrt2}&e_x&e_c\\[2pt]
\frac{e_c}{\sqrt2}&-\frac{k}{\sqrt2}&-\frac{e_c}{\sqrt2}&\frac{k}{\sqrt2}&e_c&e_p
\end{pmatrix}\left(R^{HD}\right)^T,\nonumber\\
d_{HD}=\frac{1}{\sqrt2}\,R^{HD}\begin{pmatrix}w_x\\w_p\\w_x\\w_p\\0\\0\end{pmatrix}.
\label{eq:phase-VHD}
\end{gather}

Applying the rule from equations \ref{eq:hdrule} -- \ref{eq:pimatrixhdrule} first for the X measurement and then for the P measurement, we obtain the expression describing the output mode
\begin{gather}
V^{HD''}=V^{E}-\sinh^2 2r\cos^2\phi\;R_2\,Q^{-1}R_2^T,\nonumber\\
d_{HD''}=\sinh 2r\cos\phi\;R_2\,Q^{-1}\left[R\,w-\sqrt2\begin{pmatrix}m_{xn}\\ m_{pn}\end{pmatrix}\right],\nonumber\\
Q=R\,W R^T+N,\qquad
R_2=R\,R=\begin{pmatrix}\cos2\phi&\sin2\phi\\-\sin2\phi&\cos2\phi\end{pmatrix},\qquad
V^{E}=\begin{pmatrix}e_x&e_c\\e_c&e_p\end{pmatrix},\nonumber\\
N=\begin{pmatrix}
e_x\cos^2\phi+2e_c\sin\phi\cos\phi+e_p\sin^2\phi&\left(e_x-e_p\right)\sin\phi\cos\phi-e_c\left(\cos^2\phi-\sin^2\phi\right)\\[4pt]
\left(e_x-e_p\right)\sin\phi\cos\phi-e_c\left(\cos^2\phi-\sin^2\phi\right)&e_x\sin^2\phi-2e_c\sin\phi\cos\phi+e_p\cos^2\phi
\end{pmatrix},
\label{eq:phase-afterP}
\end{gather}
where, as previously, $m_{xn}$ and $m_{pn}$ are the outcomes of the X and P measurements for teleportation $n$. The same factor is present both in the covariance matrix and in the displacement vector in equation \ref{eq:phase-afterP}. We define it as the gain of single teleportation, which is now a matrix:
\begin{equation}
T=\sinh 2r\cos\phi\;R_2\,Q^{-1}.
\label{eq:phase-T}
\end{equation}

The output mode is then displaced again by $d_{p}$, analogically to previous cases. The covariance matrix and displacement vector of the output state are
\begin{equation}
V^{OUT}=V^{E}-\sinh 2r\cos\phi\;T\,R_2^T,\qquad
d_{OUT}=T\left[R_2\,d_{IN}-\sqrt2\begin{pmatrix}m_{xn}\\ m_{pn}\end{pmatrix}\right]+\begin{pmatrix}0\\ d_{p}\end{pmatrix}.
\label{eq:phase-map}
\end{equation}
Just as in previous cases the displacement of the input state and the outcomes of the measurements are multiplied by the gain, and the newly injected displacement is added. The differences are that the input state is rotated by $2\phi$ and that the gain factor is defined as matrix. Thus, the outcomes of the X measurements now influence the results of P measurements. It is easy to check that for $\phi=0$ and $V^{IN}=\begin{pmatrix}1&0\\0&1\end{pmatrix}$ we recover the ideal case:
\begin{gather}
R_2=\begin{pmatrix}1&0\\0&1\end{pmatrix},\qquad N=V^{E}=\cosh 2r\begin{pmatrix}1&0\\0&1\end{pmatrix},\qquad Q=(1+\cosh 2r)\begin{pmatrix}1&0\\0&1\end{pmatrix},\nonumber\\
T=\frac{\sinh 2r}{1+\cosh 2r}\begin{pmatrix}1&0\\0&1\end{pmatrix}=\tanh r\begin{pmatrix}1&0\\0&1\end{pmatrix},\qquad
V^{OUT}=\left(\cosh 2r-\sinh 2r\tanh r\right)\begin{pmatrix}1&0\\0&1\end{pmatrix}=\begin{pmatrix}1&0\\0&1\end{pmatrix}.
\label{eq:phase-ideal-check}
\end{gather}

Following the experimental protocol the output state of one teleportation is the input state of the next one. We set covariance matrix of the output state after $n$ teleportation to $V_n$, its displacement vector to be the $d_n$ and the gain related to teleportation $n$ to be $T_n$. The initial probe (equation \ref{eq:probe0}) has $V_0$ of identity matrix and displacement $d_{p}$ only in P direction. From equations \ref{eq:phase-T} and \ref{eq:phase-map} we write the general formula for calculating gain of teleportation and covariance matrix and displacement vector of output state after teleportation:
\begin{gather}
Q_n=R_2\,V_{n-1}R_2^T+N,\qquad T_n=\sinh 2r\cos\phi\;R_2\,Q_n^{-1},\qquad V_n=V^{E}-\sinh 2r\cos\phi\;T_n\,R_2^T,\nonumber\\
d_n=T_n\left[R_2\,d_{n-1}-\sqrt2\begin{pmatrix}m_{xn}\\ m_{pn}\end{pmatrix}\right]+\begin{pmatrix}0\\ d_{p}\end{pmatrix} .
\label{eq:phase-rec}
\end{gather}
Already after the first teleportation covariance matrix is not the identity, thus the probe is not a coherent state and the gain changes with every teleportation. The matrices $Q_n$, $T_n$ and $V_n$ are therefore calculated recursively. The displacement vector $d_n$ depends linearly on the injected displacement $d_{p}$ and on the outcomes of the measurements. We separate these two parts,
\begin{equation}
d_n=d_{p}\,a_n+b_n ,
\label{eq:phase-ab}
\end{equation}
where the vector $a_n$ describes how the injected displacement accumulates and the vector $b_n$ contains all terms with the outcomes. Inserting this into equation \ref{eq:phase-rec} gives
\begin{gather}
a_n=T_n\,R_2\,a_{n-1}+\begin{pmatrix}0\\1\end{pmatrix},\qquad a_0=\begin{pmatrix}0\\1\end{pmatrix},\nonumber\\
b_n=T_n\left[R_2\,b_{n-1}-\sqrt2\begin{pmatrix}m_{xn}\\ m_{pn}\end{pmatrix}\right],\qquad b_0=\begin{pmatrix}0\\0\end{pmatrix}.
\label{eq:phase-ab-rec}
\end{gather}

Following the same pattern as in the ideal case for the readout of information, we derive the equation for the outcome of the final measurements, however this time including phase noise:
\begin{gather}
z\equiv\frac{m_{pfinal1}+m_{pfinal2}}{\sqrt2},\nonumber\\
\langle z\rangle=\left[R_2\,d_M\right]_p=d_{p}\left[R_2\,a_M\right]_p+\left[R_2\,b_M\right]_p,\qquad
\mathrm{Var}(z)=\left[R_2\,V_M R_2^T\right]_{pp},
\label{eq:phase-readout}
\end{gather}
where $m_{pfinal1}$, $m_{pfinal2}$ are outcomes of final measurements of homodyne one and two respectively, $[\,\cdot\,]_p$ is the second ($P$) entry of a displacement vector and $[\,\cdot\,]_{pp}$ is the lower right entry of a covariance matrix. Solving for $d_{p}$ we obtain the formula for the displacement that our input state undergoes and sensitivity of our measurements. The complete result of the final measurement, together with all variables it contains, is

\begin{equation}
\setlength{\fboxsep}{10pt}
\boxed{\begin{gathered}
\hat d_{p}=\frac{z-\left[R_2\,b_M\right]_p}{\left[R_2\,a_M\right]_p},\qquad
\sigma=\sqrt{\mathrm{Var}(\hat d_{p})}=\frac{\sqrt{\left[R_2\,V_M R_2^T\right]_{pp}}}{\left|\left[R_2\,a_M\right]_p\right|},\\[10pt]
z=\frac{m_{pfinal1}+m_{pfinal2}}{\sqrt2},\\[4pt]
a_n=T_n\,R_2\,a_{n-1}+\begin{pmatrix}0\\1\end{pmatrix},\qquad a_0=\begin{pmatrix}0\\1\end{pmatrix},\\[4pt]
b_n=T_n\left[R_2\,b_{n-1}-\sqrt2\begin{pmatrix}m_{xn}\\ m_{pn}\end{pmatrix}\right],\qquad b_0=\begin{pmatrix}0\\0\end{pmatrix},\\[4pt]
T_n=\sinh 2r\cos\phi\;R_2\,Q_n^{-1},\qquad Q_n=R_2\,V_{n-1}R_2^T+N,\\[4pt]
V_n=V^{E}-\sinh 2r\cos\phi\;T_n\,R_2^T,\qquad V_0=\begin{pmatrix}1&0\\0&1\end{pmatrix},\\[4pt]
R_2=\begin{pmatrix}\cos2\phi&\sin2\phi\\-\sin2\phi&\cos2\phi\end{pmatrix},\qquad
V^{E}=\begin{pmatrix}e_x&e_c\\e_c&e_p\end{pmatrix},\\[4pt]
N=\begin{pmatrix}
e_x\cos^2\phi+2e_c\sin\phi\cos\phi+e_p\sin^2\phi&\left(e_x-e_p\right)\sin\phi\cos\phi-e_c\left(\cos^2\phi-\sin^2\phi\right)\\[4pt]
\left(e_x-e_p\right)\sin\phi\cos\phi-e_c\left(\cos^2\phi-\sin^2\phi\right)&e_x\sin^2\phi-2e_c\sin\phi\cos\phi+e_p\cos^2\phi
\end{pmatrix},\\[4pt]
e_x=\cosh 2r-\sinh 2r\sin^2\phi,\qquad e_p=\cosh 2r+\sinh 2r\sin^2\phi,\qquad e_c=-\sinh 2r\sin\phi\cos\phi
\end{gathered}}
\label{eq:phase-estimate}
\end{equation}
Here $\hat d_{p}$ is the estimate of the injected displacement and $\sigma$ is its standard deviation (sensitivity), $M$ is the number of teleportations, $m_{xn}$ and $m_{pn}$ are the outcomes of the homodyne measurements of the $X$ and $P$ quadrature in teleportation $n$, $m_{pfinal1}$ and $m_{pfinal2}$ are the outcomes of the final measurements, $r$ is the squeezing parameter and $\phi$ is the phase shift. The quantities $Q_n$, $T_n$, $V_n$, $a_n$ and $b_n$ are calculated one after another for $n=1,\dots,M$.

\subsubsection{Model with losses and phase noise included}
\label{sec:analytical_model_losses_phase_noise}

Here we combine the two imperfections treated in sections \ref{sec:analytical_model_losses} and \ref{sec:analytical_model_phase_noise} to obtain the full
model, which describes the experiment under realistic conditions.

The phase noise is modeled as a phase shift $\phi$ acting on the relevant modes and the loss is again modeled by a fictitious beam splitter of transmission $\eta$ right before each homodyne detector. Up to the cluster beam splitter nothing changes with respect to section \ref{sec:analytical_model_phase_noise}. The input state is given by equation \ref{eq:phase-input}, the EPR state by equation \ref{eq:phase-VEPR}, the rotated input state $W$, $w$ by equation \ref{eq:phase-W} and the state after the cluster beam splitter $V^{CL}$, $d_{CL}$ is represented by equation \ref{eq:phase-VCL}. As in the case with phase noise only, the displacement in $X$ quadrature is not set to zero.

Between the cluster beam splitter and the homodyne detectors the modes being measured are rotated once more by $\phi$ and then experience loss. The covariance matrix and displacement vector of the state at the input of the detectors, together with all variables they contain, read
\begin{gather}
V^{HD}=L\,R^{HD}\,V^{CL}\left(R^{HD}\right)^TL^T+N^{loss},\qquad d_{HD}=L\,R^{HD}\,d_{CL},\nonumber\\[6pt]
L=\begin{pmatrix}\sqrt\eta&0&0&0&0&0\\0&\sqrt\eta&0&0&0&0\\0&0&\sqrt\eta&0&0&0\\0&0&0&\sqrt\eta&0&0\\0&0&0&0&1&0\\0&0&0&0&0&1\end{pmatrix},\qquad
N^{loss}=\begin{pmatrix}1-\eta&0&0&0&0&0\\0&1-\eta&0&0&0&0\\0&0&1-\eta&0&0&0\\0&0&0&1-\eta&0&0\\0&0&0&0&0&0\\0&0&0&0&0&0\end{pmatrix},\nonumber\\[6pt]
R^{HD}=\begin{pmatrix}\cos\phi&\sin\phi&0&0&0&0\\-\sin\phi&\cos\phi&0&0&0&0\\0&0&\cos\phi&\sin\phi&0&0\\0&0&-\sin\phi&\cos\phi&0&0\\0&0&0&0&1&0\\0&0&0&0&0&1\end{pmatrix},\nonumber\\[6pt]
V^{CL}=\begin{pmatrix}
\frac{W_{xx}+e_x}{2}&\frac{W_{xp}+e_c}{2}&\frac{W_{xx}-e_x}{2}&\frac{W_{xp}-e_c}{2}&\frac{k}{\sqrt2}&\frac{e_c}{\sqrt2}\\[2pt]
\frac{W_{xp}+e_c}{2}&\frac{W_{pp}+e_p}{2}&\frac{W_{xp}-e_c}{2}&\frac{W_{pp}-e_p}{2}&\frac{e_c}{\sqrt2}&-\frac{k}{\sqrt2}\\[2pt]
\frac{W_{xx}-e_x}{2}&\frac{W_{xp}-e_c}{2}&\frac{W_{xx}+e_x}{2}&\frac{W_{xp}+e_c}{2}&-\frac{k}{\sqrt2}&-\frac{e_c}{\sqrt2}\\[2pt]
\frac{W_{xp}-e_c}{2}&\frac{W_{pp}-e_p}{2}&\frac{W_{xp}+e_c}{2}&\frac{W_{pp}+e_p}{2}&-\frac{e_c}{\sqrt2}&\frac{k}{\sqrt2}\\[2pt]
\frac{k}{\sqrt2}&\frac{e_c}{\sqrt2}&-\frac{k}{\sqrt2}&-\frac{e_c}{\sqrt2}&e_x&e_c\\[2pt]
\frac{e_c}{\sqrt2}&-\frac{k}{\sqrt2}&-\frac{e_c}{\sqrt2}&\frac{k}{\sqrt2}&e_c&e_p
\end{pmatrix},\qquad
d_{CL}=\frac{1}{\sqrt2}\begin{pmatrix}w_x\\w_p\\w_x\\w_p\\0\\0\end{pmatrix},\nonumber\\[6pt]
\begin{pmatrix}W_{xx}&W_{xp}\\W_{xp}&W_{pp}\end{pmatrix}=R\,V^{IN}R^T,\qquad
\begin{pmatrix}w_x\\w_p\end{pmatrix}=R\,d_{IN},\qquad
R=\begin{pmatrix}\cos\phi&\sin\phi\\-\sin\phi&\cos\phi\end{pmatrix},\nonumber\\[6pt]
e_x=\cosh 2r-\sinh 2r\sin^2\phi,\qquad e_p=\cosh 2r+\sinh 2r\sin^2\phi,\nonumber\\
e_c=-\sinh 2r\sin\phi\cos\phi,\qquad k=\sinh 2r\cos^2\phi .
\label{eq:lossphase-apply}
\end{gather}

Similarly to the previous cases we perform now Bell measurement. Applying the rule from equations \ref{eq:hdrule} -- \ref{eq:pimatrixhdrule} first for the X measurement and then for the P measurement, we obtain the expression describing the output mode
\begin{gather}
V^{HD''}=V^{E}-\sinh^2 2r\cos^2\phi\;R_2\,Q^{-1}R_2^T,\nonumber\\
d_{HD''}=\sinh 2r\cos\phi\;R_2\,Q^{-1}\left[R\,w-\sqrt{\frac{2}{\eta}}\begin{pmatrix}m_{xn}\\ m_{pn}\end{pmatrix}\right],\nonumber\\
Q=R\,W R^T+N+2\,\frac{1-\eta}{\eta}\begin{pmatrix}1&0\\0&1\end{pmatrix},\qquad
R_2=R\,R=\begin{pmatrix}\cos2\phi&\sin2\phi\\-\sin2\phi&\cos2\phi\end{pmatrix},\qquad
V^{E}=\begin{pmatrix}e_x&e_c\\e_c&e_p\end{pmatrix},\nonumber\\
N=\begin{pmatrix}
e_x\cos^2\phi+2e_c\sin\phi\cos\phi+e_p\sin^2\phi&\left(e_x-e_p\right)\sin\phi\cos\phi-e_c\left(\cos^2\phi-\sin^2\phi\right)\\[4pt]
\left(e_x-e_p\right)\sin\phi\cos\phi-e_c\left(\cos^2\phi-\sin^2\phi\right)&e_x\sin^2\phi-2e_c\sin\phi\cos\phi+e_p\cos^2\phi
\end{pmatrix},
\label{eq:lossphase-afterP}
\end{gather}
where, as previously, $m_{xn}$ and $m_{pn}$ are the outcomes of the X and P measurements for teleportation $n$. The matrices $R_2$, $V^{E}$ and $N$ are the same as in equation \ref{eq:phase-afterP}. Compared with equation \ref{eq:phase-afterP} the loss changes two things: the term $2\frac{1-\eta}{\eta}$ is added to $Q$ and the outcomes are multiplied by $\sqrt{2/\eta}$ instead of $\sqrt2$. The same two changes appear between the ideal case and the case with losses (equations \ref{eq:loss-g} and \ref{eq:loss-map}). The gain of single teleportation has the same form as in equation \ref{eq:phase-T},
\begin{equation}
T=\sinh 2r\cos\phi\;R_2\,Q^{-1},
\label{eq:lossphase-T}
\end{equation}
but now with $Q$ from equation \ref{eq:lossphase-afterP}.

The output mode is then displaced again by $d_{p}$, analogically to previous cases. The covariance matrix and displacement vector of the output state are
\begin{gather}
V^{OUT}=V^{E}-\sinh 2r\cos\phi\;T\,R_2^T,\qquad
d_{OUT}=T\left[R_2\,d_{IN}-\sqrt{\frac{2}{\eta}}\begin{pmatrix}m_{xn}\\ m_{pn}\end{pmatrix}\right]+\begin{pmatrix}0\\ d_{p}\end{pmatrix},\nonumber\\
Q=R_2\,V^{IN}R_2^T+N+2\,\frac{1-\eta}{\eta}\begin{pmatrix}1&0\\0&1\end{pmatrix},
\label{eq:lossphase-map}
\end{gather}
where we used $W=R\,V^{IN}R^T$ and $w=R\,d_{IN}$ from equation \ref{eq:phase-W}. Following the experimental protocol the output state of one teleportation is the input state of the next one. As in section \ref{sec:analytical_model_phase_noise} we set covariance matrix of the output state after $n$ teleportation to $V_n$, its displacement vector to be the $d_n$ and the gain related to teleportation $n$ to be $T_n$. The initial probe is coherent state represented by an identity matrix with displacement $d_{p}$ only in P direction. From equations \ref{eq:lossphase-T} and \ref{eq:lossphase-map} we write the general formula for calculating gain of teleportation and covariance matrix and displacement vector of output state after teleportation:
\begin{gather}
Q_n=R_2\,V_{n-1}R_2^T+N+2\,\frac{1-\eta}{\eta}\begin{pmatrix}1&0\\0&1\end{pmatrix},\qquad T_n=\sinh 2r\cos\phi\;R_2\,Q_n^{-1},\nonumber\\
V_n=V^{E}-\sinh 2r\cos\phi\;T_n\,R_2^T,\qquad
d_n=T_n\left[R_2\,d_{n-1}-\sqrt{\frac{2}{\eta}}\begin{pmatrix}m_{xn}\\ m_{pn}\end{pmatrix}\right]+\begin{pmatrix}0\\ d_{p}\end{pmatrix} .
\label{eq:lossphase-rec}
\end{gather}
The matrices $Q_n$, $T_n$ and $V_n$ are calculated recursively. As in equation \ref{eq:phase-ab} we separate the part of $d_n$ which depends on the injected displacement $d_{p}$ from the part which depends on the outcomes of the measurements:
\begin{gather}
d_n=d_{p}\,a_n+b_n ,\nonumber\\
a_n=T_n\,R_2\,a_{n-1}+\begin{pmatrix}0\\1\end{pmatrix},\qquad a_0=\begin{pmatrix}0\\1\end{pmatrix},\nonumber\\
b_n=T_n\left[R_2\,b_{n-1}-\sqrt{\frac{2}{\eta}}\begin{pmatrix}m_{xn}\\ m_{pn}\end{pmatrix}\right],\qquad b_0=\begin{pmatrix}0\\0\end{pmatrix}.
\label{eq:lossphase-ab-rec}
\end{gather}

Following the same pattern as in the ideal case for the readout of information, we derive the equation for the outcome of the final measurements, this time including both losses and phase noise:
\begin{gather}
z\equiv\frac{m_{pfinal1}+m_{pfinal2}}{\sqrt{2\eta}},\nonumber\\
\langle z\rangle=\left[R_2\,d_M\right]_p=d_{p}\left[R_2\,a_M\right]_p+\left[R_2\,b_M\right]_p,\qquad
\mathrm{Var}(z)=\left[R_2\,V_M R_2^T\right]_{pp}+\frac{1-\eta}{\eta},
\label{eq:lossphase-readout}
\end{gather}
where $m_{pfinal1}$, $m_{pfinal2}$ are outcomes of final measurements of homodyne one and two respectively, $[\,\cdot\,]_p$ is the second ($P$) entry of a displacement vector, $[\,\cdot\,]_{pp}$ is the lower right entry of a covariance matrix and $\frac{1-\eta}{\eta}$ is the extra noise from the loss in front of the detectors. Solving for $d_{p}$ we obtain the formula for the displacement that our input state undergoes and sensitivity of our measurements. The complete result of the final measurement, together with all variables it contains, is
\begin{equation}
\setlength{\fboxsep}{10pt}
\boxed{\begin{gathered}
\hat d_{p}=\frac{z-\left[R_2\,b_M\right]_p}{\left[R_2\,a_M\right]_p},\qquad
\sigma=\sqrt{\mathrm{Var}(\hat d_{p})}=\frac{\sqrt{\left[R_2\,V_M R_2^T\right]_{pp}+\frac{1-\eta}{\eta}}}{\left|\left[R_2\,a_M\right]_p\right|},\\[10pt]
z=\frac{m_{pfinal1}+m_{pfinal2}}{\sqrt{2\eta}},\\[4pt]
a_n=T_n\,R_2\,a_{n-1}+\begin{pmatrix}0\\1\end{pmatrix},\qquad a_0=\begin{pmatrix}0\\1\end{pmatrix},\\[4pt]
b_n=T_n\left[R_2\,b_{n-1}-\sqrt{\frac{2}{\eta}}\begin{pmatrix}m_{xn}\\ m_{pn}\end{pmatrix}\right],\qquad b_0=\begin{pmatrix}0\\0\end{pmatrix},\\[4pt]
T_n=\sinh 2r\cos\phi\;R_2\,Q_n^{-1},\qquad Q_n=R_2\,V_{n-1}R_2^T+N+2\,\frac{1-\eta}{\eta}\begin{pmatrix}1&0\\0&1\end{pmatrix},\\[4pt]
V_n=V^{E}-\sinh 2r\cos\phi\;T_n\,R_2^T,\qquad V_0=\begin{pmatrix}1&0\\0&1\end{pmatrix},\\[4pt]
R_2=\begin{pmatrix}\cos2\phi&\sin2\phi\\-\sin2\phi&\cos2\phi\end{pmatrix},\qquad
V^{E}=\begin{pmatrix}e_x&e_c\\e_c&e_p\end{pmatrix},\\[4pt]
N=\begin{pmatrix}
e_x\cos^2\phi+2e_c\sin\phi\cos\phi+e_p\sin^2\phi&\left(e_x-e_p\right)\sin\phi\cos\phi-e_c\left(\cos^2\phi-\sin^2\phi\right)\\[4pt]
\left(e_x-e_p\right)\sin\phi\cos\phi-e_c\left(\cos^2\phi-\sin^2\phi\right)&e_x\sin^2\phi-2e_c\sin\phi\cos\phi+e_p\cos^2\phi
\end{pmatrix},\\[4pt]
e_x=\cosh 2r-\sinh 2r\sin^2\phi,\qquad e_p=\cosh 2r+\sinh 2r\sin^2\phi,\qquad e_c=-\sinh 2r\sin\phi\cos\phi
\end{gathered}}
\label{eq:lossphase-estimate}
\end{equation}
Here $\hat d_{p}$ is the estimate of the injected displacement and $\sigma$ is its standard deviation (sensitivity), $M$ is the number of teleportations, $m_{xn}$ and $m_{pn}$ are the outcomes of the homodyne measurements of the $X$ and $P$ quadrature in teleportation $n$, $m_{pfinal1}$ and $m_{pfinal2}$ are the outcomes of the final measurements, $r$ is the squeezing parameter, $\phi$ is the phase shift and $\eta$ is the transmission of the fictitious beam splitter in front of each homodyne detector. The quantities $Q_n$, $T_n$, $V_n$, $a_n$ and $b_n$ are calculated one after another for $n=1,\dots,M$.

\subsection{Calibration and linearity of displacement beam}
\label{sec: disp_calibration}

\begin{figure}[H]
  \centering
  \makebox[0.5\textwidth][l]{\textbf{(a)}}\\[-0.5ex]
  \includegraphics[width=0.5\textwidth]{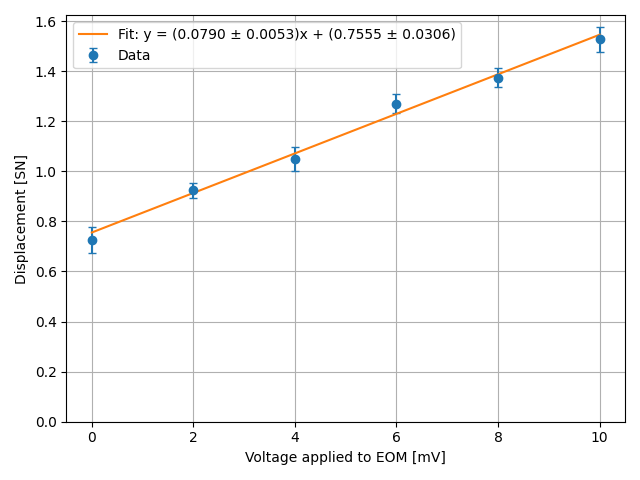}\\[0.5ex]
  \makebox[0.5\textwidth][l]{\textbf{(b)}}\\[-0.5ex]
  \includegraphics[width=0.5\textwidth]{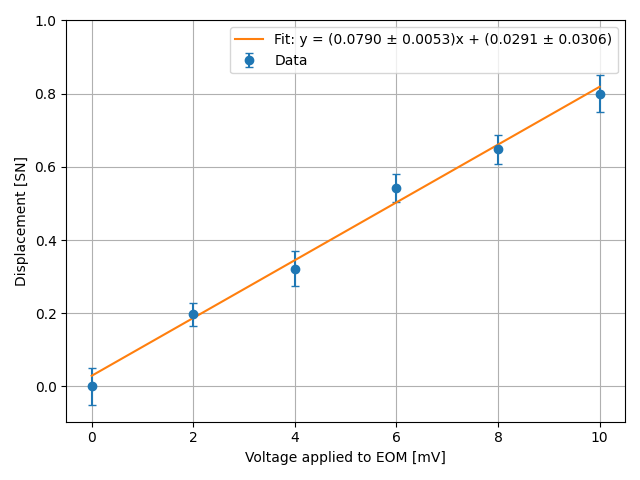}
  \caption{Calibration of the displacement. Measured displacement as a
  function of the voltage applied to the phase modulator in the displacement
  path, (a) as measured and (b) after subtraction of the static offset.
  Lines: linear fits.}
  \label{fig:disp_calibration}
\end{figure}

The displacement is calibrated before the experiment with a vacuum probe, in
the same configuration as the experiment itself, i.e.\ with all relative
phases locked. The displacement is varied by changing the voltage applied to
the phase modulator in the displacement path. Only small displacements are
used, because the phase noise of the displacement beam increases the variance
of the measurement for larger displacements, even in a single-pass
measurement. The result is shown in Fig.~\ref{fig:disp_calibration}a. The
measured displacement depends linearly on the applied voltage.

The fit also reveals an offset of $(0.76 \pm 0.03)$~SN at zero voltage, where
no displacement is expected. We find that this offset is static and appears
only when the displacement modulation and the switching of the measurement
quadrature are applied at the same time. It is stable over time and reproducible: for a given optical power it has the same magnitude each time the setup is switched on, including on
different days. It can therefore be removed by subtracting the value measured at zero voltage from all measurements, taking the sign convention of the beam splitters into account.
Figure~\ref{fig:disp_calibration}b shows the calibration data after this
correction. The slope is unchanged and the remaining offset,
$(0.03 \pm 0.03)$~SN, is consistent with zero.

The calibrated displacement used as the reference value in
Fig.~\ref{fig:results}a is obtained by multiplying the slope by the voltage
applied to the phase modulator during the experiment, and its uncertainty
follows from the uncertainty of the slope.

\subsection{Displacement vs phase sensing with cluster state based teleportation}
\label{sec: disp_vs_phase}

In the original proposal~\cite{Johanness}, the measured quantity is a phase
shift. That protocol includes a fast feedforward loop, which corrects the
displacement introduced when one mode of the cluster state is measured. This
feedforward is not implemented in our experiment, and without it the protocol
cannot be used for phase sensing. The reason is that the displacement
introduced by each measurement is random and depends strongly on the
squeezing. The higher the squeezing, the larger the displacement. In a
phase-sensing protocol this displacement is, in addition, rotated by the
phase shift. It is in principle possible to correct for the rotated
displacement in post-processing, but the correction is limited by the $\pi$
symmetry of the phase. In practice, the random displacement can move the
state into a different quadrant of phase space, in which case the direction
of the phase shift can no longer be distinguished. In the critical case, all
contributions cancel and no phase shift is detected. The problem disappears
if the displacement is corrected in real time, faster than length of one temporal
mode, or if the measured quantity is defined as a displacement along a fixed
direction in phase space. We follow the second approach.

\end{document}